\documentclass[traditabstract]{aa} 

\usepackage{amsmath}

\usepackage{graphicx}
\usepackage{txfonts}
\usepackage{natbib}
\usepackage{longtable}
\newcommand{\fluxunits}{{\rm erg}\;{\rm sec}^{-1}{\rm cm}^{-2}}

\newcommand{\alphaox}{\alpha_{\rm ox}}
\newcommand{\alphao}{\alpha_{\rm opt}}
\newcommand{\lo}{\rm L_{2500 \;\text{\AA}}}
\newcommand{\lx}{\rm L_{2 \;{\rm keV}}}
\newcommand{\z}{{z}}
\newcommand{\MC}{\text{1312-{\rm XMM}}}

\newcommand{\MBH}{M_{\rm BH}}
\newcommand{\lambdaEdd}{\lambda_{\rm Edd}}
\newcommand{\kbol}{k_{\rm bol}}
\newcommand{\Lbol}{\rm L_{\rm bol}}
\newcommand{\Lbolmu}{\rm L_{\rm bol,1\mu m}}
\newcommand{\Lhard}{\rm{L_{[2-10]keV}}}
\newcommand{\LEdd}{\rm L_{\rm Edd}}
\newcommand{\Log}{\rm Log~}
\newcommand{\Alambda}{\rm 2500 \;\text{\AA}}

\newcommand{\rev}[1]{{ #1}}

\begin{document}
   \title{The X-ray to optical-UV luminosity ratio of X-ray selected Type 1 AGN in XMM-COSMOS}

   
   \author{E. Lusso$^{1,2}$\thanks{elisabeta.lusso2@unibo.it},
          A. Comastri$^{2}$, C. Vignali$^{1,2}$, G. Zamorani$^{2}$, M. Brusa$^{3}$, R. Gilli$^{2}$, K. Iwasawa$^{2}$, M. Salvato$^{4}$, F. Civano$^{5}$, M. Elvis$^{5}$, A. Merloni$^{3,6}$, A. Bongiorno$^{3}$, J.R. Trump$^{7}$, A.M. Koekemoer$^{8}$, E. Schinnerer$^{9}$, E. Le Floc'h$^{10}$, N. Cappelluti$^{3}$, K. Jahnke$^{9}$, M. Sargent$^{11}$, J. Silverman$^{11}$, V. Mainieri$^{3,12}$, F. Fiore$^{13}$, M. Bolzonella$^{2}$, O. Le F\`{e}vre$^{14}$, B. Garilli$^{15}$, A. Iovino$^{16}$, J.P. Kneib$^{14}$, F. Lamareille$^{17}$, S. Lilly$^{11}$, M. Mignoli$^{2}$, M. Scodeggio$^{15}$, D. Vergani$^{1,2}$.
       }

   \institute{$^{1}$Dipartimento di Astronomia, Universit\`{a} di Bologna, via Ranzani 1, I-40127 Bologna, Italy.\\
$^{2}$INAF-Osservatorio Astronomico di Bologna, via Ranzani 1, I-40127 Bologna, Italy.\\
$^{3}$Max Planck Institut f\"{u}r extraterrestische Physik, Giessenbachstrasse 1, 85748 Garching, Germany.\\
$^{4}$California Institute of Technology, MC 105-24, 1200 East California Boulevard, Pasadena, CA 91125, USA.\\
$^{5}$Harvard-Smithsonian Center for Astrophysics, 60 Garden Street,Cambridge, MA 02138, USA.\\
$^{6}$Excellence Cluster Universe, Boltzmannstr. 2, 85748, Garching, Germany. \\
$^{7}$Steward Observatory, University of Arizona, 933 N Cherry Ave, Tucson, AZ 85721, USA.\\
$^{8}$Space Telescope Science Institute, 3700 San Martin Drive, Baltimore, MD 21218. \\
$^{9}$Max Planck Institut f\"{u}r Astronomie, K\"{o}nigstuhl 17, D-69117 Heidelberg, Germany.\\
$^{10}$Institute for Astronomy, University of Hawaii, 2680 Woodlawn Drive, Honolulu, HI 96822-1839, USA. \\
$^{11}$Department of Physics, ETH Zurich, CH-8093 Zurich, Switzerland. \\
$^{12}$ESO, Karl-Schwarschild-Strasse 2, DY85748 Garching, Germany. \\
$^{13}$INAF–Osservatorio Astronomico di Roma, via Frascati 33, Monteporzio (Rm), I00040, Italy. \\
$^{14}$Laboratoire d’Astrophysique de Marseille, OAMP, CNRS-Universit\'{e} Aix-Marseille, 38, rue Fr\'{e}d\'{e}ric Joliot-Curie, 13388 Marseille cedex 13, France. \\
$^{15}$INAF--IASF, via Bassini 15, 20133 Milano, Italy. \\
$^{16}$INAF--Osservatorio Astronomico di Brera, via Brera 28, 20121 Milan, Italy. \\
$^{17}$Laboratoire d’Astrophysique de Toulouse-Tarbes, Universit\'{e} de Toulouse, CNRS, 14 avenue Edouard Belin, F-31400 Toulouse, France. \\
}

\authorrunning{E. Lusso et al}

   \date{Revised version, \today} 

  \abstract{We present a study of the X-ray to optical properties of a sample of 545 X-ray selected Type 1 AGN, from the XMM-COSMOS survey, over a wide range of redshifts ($0.04<\z<4.25$) and X-ray luminosities ($40.6 \leq \Log \Lhard \leq 45.3$). About 60\% of them are spectroscopically identified Type 1 AGN, while the others have a reliable photometric redshift and are classified as Type 1 AGN on the basis of their multi-band Spectral Energy Distributions. We discuss the relationship between UV and X-ray luminosity, as parameterized by the $\alphaox$ spectral slope, and its dependence on redshift and luminosity. We compare our findings with previous investigations of optically selected broad-line AGN (mostly from SDSS). A highly significant correlation between $\alphaox$ and $\lo$ is found, in agreement with previous investigations of optically selected samples.
We calculate bolometric corrections, $\kbol$, for the whole sample using hard X-ray luminosities ($\Lhard$), and the Eddington ratios for a subsample of 150 objects for which black hole mass estimates are available. We confirm the trend of increasing bolometric correction with increasing Eddington ratio as proposed in previous works. A tight correlation is found between $\alphaox$ and $\kbol$, which can be used to estimate accurate bolometric corrections using only optical and X-ray data.  We find a significant correlation between $\alphaox$ and Eddington ratio, in which $\alphaox$ increases for increasing Eddington ratios.}
   \maketitle
%

\section{Introduction}

The distribution of X-ray to optical--UV  ratio in quasars, as a function of optical and X-ray luminosity and redshift, has been the subject of active investigations for more than 30 years (\citealt{avnitananbaum79}).
The ratio is parameterized by the optical to X-ray spectral index ($\alphaox$) defined as:
\begin{equation}
\label{aoxdef}
\alphaox=-\frac{\Log[\lx/\lo]}{2.605}.
\end{equation}
In the past, it was widely adopted to compute the QSO contribution to the X-ray background and estimate the X-ray luminosity function from optical counts (e.g., \citealt{zamorani81}; \citealt{avnitananbaum86}, \citealt{2005ApJ...618..123S}). 
These studies, based on {\it Einstein} observations, found a mean value of $\alphaox$ in the range  $1.3-1.45$. 
Subsequently, many investigations have been performed (e.g., \citealt{yuansiebertbrink98}; \citealt{vignali03}; \citealt{strateva05}; \citealt{steffen06}, hereafter S06; 
\citealt{just07}, hereafter J07; \citealt{kelly08}; \citealt{green09} and \citealt{2009ApJS..183...17Y}) which were mainly based on AGN samples selected from the Sloan Digital Sky Survey (SDSS). Most of the previous studies on the relationship between $\alphaox$, luminosity and redshift were based on large samples of optically selected type 1 AGN with a high fraction of X-ray detections. The S06 paper extends the \citet{strateva05} work including moderate-luminosity optically selected AGN from the COMBO-17 survey with deep observations in the \textit{Chandra} Deep Field-South. Their full sample consists of 333 optically selected Type 1 AGN with a high fraction of X-ray detections ($\sim 88\%$) in the redshift range $0.01\leq\z\leq6.28$. More recently, J07 have further extended the S06 sample with 59 optically luminous quasars spanning a redshift range of 1.5--4.6 and 14 quasars from \citet{shemmer06}.
All the X-ray information on their sample is derived from \textit{Chandra}, XMM-\textit{Newton} and ROSAT observations.
The X-ray to optical flux ratio distribution of an X-ray selected sample of 188 Type 1 AGN from the ChaMP survey, which is limited to bright X-ray fluxes and optical luminosities ($f_x > 10^{14} \,\rm{erg \,cm^{-2} s^{-1}}$ and $r < 22.5$), is presented in \citet{2005ApJ...618..123S}.
A strong correlation between $\alphaox$ and the optical luminosity at 2500$\text{\AA}$ is found, while $\alphaox$ is only marginally dependent upon redshift (but see \citealt{bechtold03} for different results). The $\alphaox$ distributions typically cover the range 1.2--1.8, with a mean value of about 1.5. 

These studies could not properly address the effect of a source selection in a different band (i.e. X-rays).
The availability of large samples of X-ray selected QSOs with a high quality photometric and spectroscopic coverage in the optical now opens the possibility for an extended investigation of the $\alphaox$ distribution and its evolution. 

Understanding how $\alphaox$ evolves with luminosity and redshift may provide a first hint about the nature of the energy generation mechanism in AGN. It is also a first step towards an estimate of the AGN bolometric luminosity function (\citealt{hopkins07}) and the mass function of Supermassive Black Holes (SMBHs) (e.g. \citealt{marconi04}), and towards the understanding of the structure of the AGN accretion disk and X-ray corona. While the calculation of the bolometric luminosity requires good-quality data over a large portion of the electromagnetic spectrum, it is relatively easy to compute $\alphaox$ for sizable samples of objects up to high redshift.

We analyze the dependence of $\alphaox$ upon redshift, optical and X-ray luminosities using a large X-ray selected sample of Type 1 AGN in the Cosmic Evolution Survey (COSMOS) field (\citealt{scoville07}).
The COSMOS field is a so far unique area for deep and wide comprehensive multi-wavelength coverage: radio with the VLA, infrared with {\it Spitzer}, optical bands with {\it Hubble}, {\it Subaru}, SDSS and other ground-based telescopes, near- and far-ultraviolet bands with the \textit{Galaxy Evolution Explorer} (GALEX) and X-rays with XMM--{\it Newton} and {\it Chandra}. The spectroscopic coverage with VIMOS/VLT and IMACS/Magellan, coupled with the reliable photometric redshifts derived from multiband fitting, allows us to build a large and homogeneous sample of QSOs with a well sampled spectral coverage and to keep selection effects under control. 

The broad-band information contained in the COSMOS database is well suited for a detailed study of AGN Spectral Energy Distributions (SEDs), bolometric luminosities ($\Lbol$) and bolometric corrections, in particular the one from the X-rays, defined as:
\begin{equation}
 \kbol=\Lbol/\Lhard.
\end{equation}
If the black hole mass is available ($\MBH$), it is also possible to compute the Eddington ratio ($\lambdaEdd=\Lbol/\LEdd$, where $\LEdd=1.38 \times 
10^{38} \MBH/M_{\odot}$) and investigate any possible dependence of $\lambdaEdd$ on the SED shape. 
In particular, it is possible to test whether the correlation between hard X-ray bolometric correction $\kbol$  and Eddington ratio $\lambdaEdd$, recently obtained for bright AGN in the nearby Universe (\citealt{vasudevanfabian09}; hereafter VF09), holds also at higher redshifts. 

This paper is organized as follows. In Section \ref{The Data Set} we describe the selection criteria for the sample used in this work. Section \ref{Rest-frame monochromatic fluxes and Spectral Energy Distributions} presents the data and the method by which we construct SEDs. Data analysis and results are given in Section \ref{Statistical Analysis}. In Section \ref{Effects of reddening and host galaxy light} we estimate the possible effects of reddening and host-galaxy light contribution on our main results. The discussion of our findings is given in Section \ref{Discussion} and a summary of the results is given in Section \ref{Summary and conclusions}.
We adopted a flat model of the universe with a Hubble constant $H_{0}=70\, \rm{km \,sec^{-1}\, Mpc^{-1}}$, $\Omega_{M}=0.29$, $\Omega_{\Lambda}=1-\Omega_{M}$ (\citealt{komatsu09}).

\begin{table}[ct]
\caption{Selection criteria from $\MC$ catalog. \label{tbl-0}}
\begin{tabular}{rccl}
\hline\hline\noalign{\smallskip}
    Catalog    &  BL$^{\mathrm{a}}$   &  Photoz   &  Total$^{\mathrm{b}}$          \\

\hline\hline\noalign{\smallskip}
    1312XMM$^{\mathrm{c}}$  &  361    &  613$^*$    &     974         \\
\hline\noalign{\smallskip}
Class $\geq 19$ $^{\mathrm{d}}$  &    (361)$^{\mathrm{f}}$    &  236        &   597   \\
\hline\noalign{\smallskip}
Radio quiet$ ^{\mathrm{e}}$  & 322 &    223     &   545         \\
\hline\hline\noalign{\smallskip}
\end{tabular}
\begin{list}{}{}
\item[$^{\mathrm{a}}$]Broad-line sources from optical spectroscopy (FWHM $>$2000 km s$^{-1}$).
\item[$^{\mathrm{b}}$] This column gives the sum of the objects with spectroscopic or photometric redshifts.
\item[$^{\mathrm{c}}$] $F_{[0.5-2]{\rm keV}}  \geq 10^{-15}\,\fluxunits$ and secure optical association. The $\MC$ catalog also comprises: 241 sources which are not broad-line AGN; 34 sources which are spectroscopically classified as stars; 49 sources which are best-fitted with a stellar template and 14 sources for which the photometric redshift is not reliable.
\item[$^{\mathrm{d}}$]According to Table 2 in \citet{salvato09}.
\item[$^{\mathrm{e}}$]$R \leq 10$. Radioloudness is defined in eq. (\ref{radioloudness}). We exclude 51 radio-loud sources: 39 with spectroscopic redshift and 12 with photometric redshift.
\item[$^{\mathrm{f}}$]All spectroscopically identified broad-line AGN are included, even if the classification is less than 19.
\item[$^*$]Only extragalactic sources are considered.
\end{list}
\end{table}

\section{The Data Set}
\label{The Data Set}


\subsection{The Parent Sample}
\label{The Parent Sample}

The XMM-COSMOS catalog comprises $1822$ point--like X-ray sources detected by XMM-\textit{Newton} over an area of $\sim 2\,\rm deg^2$ for a total of $\sim 1.5$ Ms at a fairly homogeneous depth of $\sim 50$ ks (\citealt{hasinger07}, \citealt{cappelluti09}).
Following \citet{brusa09}, we excluded 24 sources which turned out to be a blend of two {\it Chandra} sources leading to a total of 1798 X-ray selected point-like sources.

Spectroscopic redshifts for the proposed counterparts are compiled by \citet{brusa09} from the Magellan/IMACS and MMT observation campaigns ($\sim 590$ objects, \citealt{trump09}), from the zCOSMOS project ($\sim 350$ objects, \citealt{lilly07}), or were already present either in the SDSS survey catalog ($\sim 100$ objects, \citealt{adelmanmccarthy05}, \citealt{kauffmann03}\footnote{These sources have been retrieved from the Nasa Extragalactic Database (NED) and from the SDSS archive.}), or in the literature ($\sim 95$ objects, \citealt{prescott06}). In summary, good-quality spectroscopic redshifts are available for 738 sources, corresponding to a substantial fraction ($\sim 45$\%) of the entire XMM--\textit{Newton} sample.

Photometric redshifts for almost all XMM--COSMOS sources have been obtained exploiting the COSMOS multi-wavelength database and are presented in Salvato et al. (2009, hereafter S09). Since the large majority of the XMM--COSMOS sources are AGN, in addition to the standard photometric redshift treatments for normal galaxies, a new set of SED templates has been adopted, together with a correction for long--term variability and  luminosity priors for point-like sources (see below and S09 for further details). The availability of the intermediate band {\it Subaru} filters (\citealt{taniguchi07}) is crucial in picking up emission lines (see also \citealt{wolf04}). This led, for the first time for an AGN sample, to a photometric redshift  accuracy comparable to that achieved for inactive galaxies ($\sigma_{\Delta \z/(1+\z)} \sim 0.015$ and $\sim 4\%$ outliers) down to $i\simeq$22.5. At fainter magnitudes (22.5 $<i<24.5$), the dispersion increases to $\sigma_{\Delta \z/(1+\z)} \simeq 0.035$ with $\sim4.8\%$ outliers, still remarkably good for an AGN sample.
A photometric redshift is available for all but 32 objects out of 1798.

In addition to the photometric redshifts, S09 provide also a photometric classification based on the best-fit broad-band SED template. Briefly, each AGN SED has been fitted with a total of 30 different templates which include both normal galaxies (early type, late type and ULIRG galaxies), low-and-high luminosity QSOs (both Type 1 and 2) and hybrids created assuming a varying ratio between the AGN and a galaxy templates (90:10, 80:20,..., 10:90; see S09 for details and \citealt{polletta07}). About 40\% of the sources are best-fitted by AGN-dominated SED, while the remaining sources are reproduced by host galaxy-dominated SED. The photometric classification is also confirmed a posteriori (see Fig.10 in S09) with the distribution of the XMM sources in the X-ray hardness ratio plane (\citealt{cappelluti09}; \citealt{hasinger07}).

We restricted the analysis to the X-ray sources detected in the soft band at a flux larger than $10^{-15}\fluxunits$ (\citealt{cappelluti09}). Given that objects for which no secure optical counterpart could be assigned, are usually affected by severe blending problems, making the photo-z estimate  unreliable, our parent sample consists of 1312 sources (hereafter $\MC$) for which a secure optical counterpart can be associated (see discussion in \citealt{brusa09,brusa09b}).

\subsection{Type 1 AGN Sample}
\label{The Broad Line AGN Sample}

From the $\MC$ catalog we have selected $361$ spectroscopically classified broad-line AGN\footnote{The origin of spectroscopic redshifts for the $361$ sources is as follows: $63$ objects from the SDSS archive, $75$ from MMT observations (\citealt{prescott06}), $112$ from the IMACS observation campaign (\citealt{trump07}), $93$ from the zCOSMOS bright $10$k sample (see \citealt{lilly07}) and $18$ from the zCOSMOS faint catalog.} on the basis of broad emission lines ($FWHM > 2000 \,{\rm km \; sec^{-1}}$) in their optical spectra. We will refer to this sample as the ``spectro-z" sample. As a comparison, in the $\MC$ catalog there are 241 objects spectroscopically classified as not broad-line AGN (Type 2 or emission-line or absorption-line galaxies).

\begin{figure}
 \resizebox{\hsize}{!}{\includegraphics{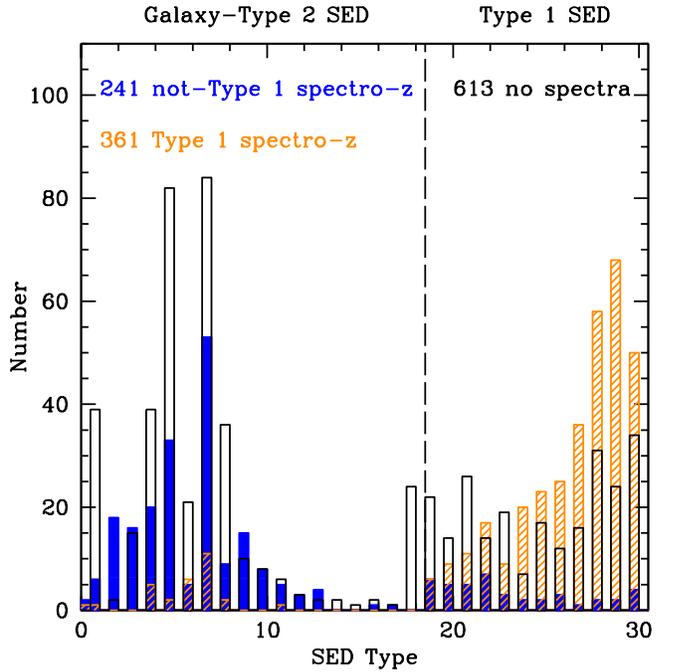}}
 \caption{Distribution of the photometric classification (see Table 2 in \citealt{salvato09}) for the spectroscopically identified Type 1 AGN sample (\textit{hatched histogram}), for the photometric sample (\textit{open histogram}) and for the not-Type 1 AGN sample (\textit{filled histogram}). The dashed line at SED Type=19 marks our adopted separation between sources with photometric redshift that we define as Type 1 AGN ($236$ sources on the right) and not-Type 1 ($377$ sources on the left) on the basis of the best-fitting SED template.}
 \label{hist1phot}
\end{figure}

\begin{figure}
 \resizebox{\hsize}{!}{\includegraphics{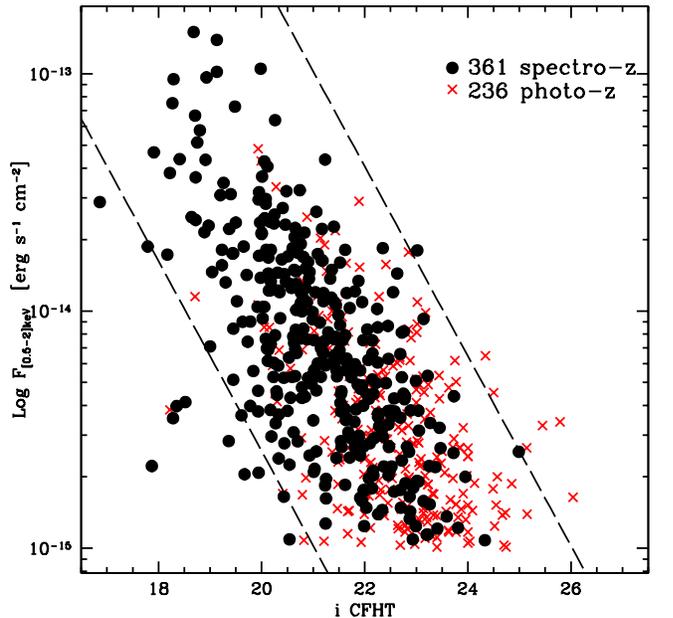}}
 \caption{Plot of the $[0.5-2]$keV flux versus the total $i^*$ CFHT magnitude for the spectroscopically identified (\textit{black points}) and the photo-z classified (\textit{red crosses}) broad line AGN. The dashed lines represent the constant X-ray to optical flux ratio $\log(X/O)=\pm 1$.} 
 \label{fluxSi}
\end{figure}

\begin{figure}
 \resizebox{\hsize}{!}{\includegraphics{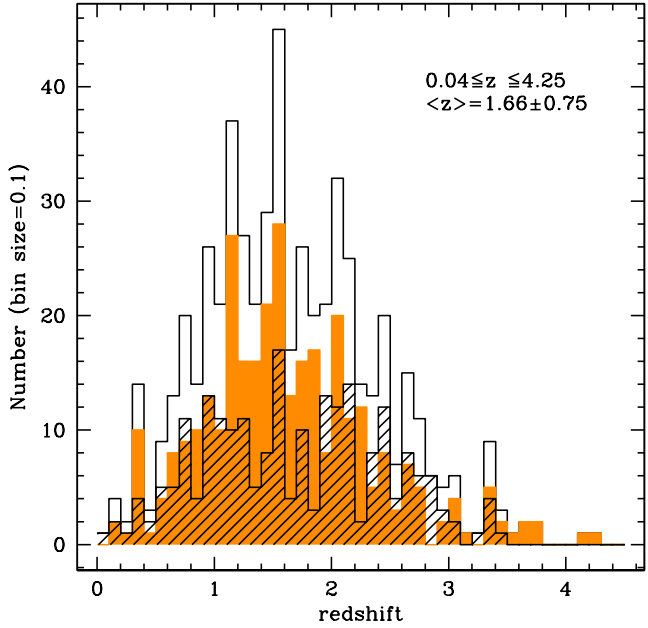}}
 \caption{Redshift distribution of the $545$ Type 1 AGN considered in this work (\textit{open histogram}). The filled histogram shows the redshift distribution for the sample of $322$ spectroscopically identified sources, while the hatched histogram is the redshift distribution for the $223$ sources without spectroscopic redshift.}
 \label{histz}
\end{figure}


The spectroscopic completeness of the X-ray selected sample quickly decreases towards faint optical magnitudes. A sizable fraction of Type 1 AGN may be present among optically faint X-ray sources; not to use them in our analysis would significantly bias the distribution of X-ray to optical flux ratios. In order to extend our Type 1 AGN sample to fainter magnitude, we proceed as follows.

First, we looked at the distribution of the best-fit photometric classifications for the spectroscopically identified sample. Figure \ref{hist1phot} shows the distribution separately for the 361 spectroscopically identified Type 1 AGN (\textit{hatched histogram}) and for the 241 sources which are not broad-line AGN (\textit{filled histogram}). The dashed line at the SED Type=19 marks the division between galaxy-dominated and AGN-dominated SED (see Table 2 in S09 for details). The large majority of the broad emission line AGN in the spectro-z sample ($>$ 90\%) are classified as Type 1 AGN by the SED fitting.
The distribution of the 241 not broad-line sources is in agreement with the SED-based classification ($\sim$ 83\% have Class$<$19), and the number of not-Type 1 which have Class$\geq$19 is relatively small (42 sources, about 17\%).
Then, we consider the remaining 613 X-ray sources in the $\MC$ sample for which only a photometric redshift is available from S09, and exclude all the sources which are best-fitted with a stellar template.
The distribution of the best-fit templates for this photometric sample is also shown in Fig. \ref{hist1phot} (\textit{open histogram}). In the following, we assume that the 236 X-ray sources, classified by the SED fitting with an AGN-dominated SED are Type 1 AGN. We will refer to this sample as the ``photo-z" sample.

In Figure \ref{fluxSi} we plot the soft X-ray flux as a function of $i^*$ CFHT magnitude. Points and crosses represent sources in the spectro-z and photo-z sample, respectively. The dashed lines delimit the region typically occupied by AGN along the X-ray to optical flux ratio $\log(X/O)=\pm1$\footnote{$\log(X/O)=\log f_x+i^*/2.5+5.6$}. The 236 sources in the photo-z sample increase the completeness of the data-set at fainter magnitudes. While some Type 1 AGN may have been missed (about $10\%$, with SED Type$<$19 in Fig. \ref{hist1phot}), we are confident that the described choice minimizes the selection bias against optically faint Type 1 AGN.
We only include sources that have optical-UV data in order to compute monochromatic luminosities at $2500\textrm{\small\AA}$ and the SED. This selection criterium only excludes the photometric source with XID=5120.

It is well known that radio-loud AGN have an enhanced X-ray emission mechanism linked to the jets, which can provide an increment in the X-ray emission with respect to radio-quiet AGN with similar optical luminosities (e.g., \citealt{zamorani81}, \citealt{wilkeselvis87rl}, \citealt{cappi97}). We exclude RL AGN from the total sample using the ``standard" definition of radio-loudness, $\rm R \geq 10$ (\citealt{kellermann89}).\footnote{Radio-loudness is defined as
\begin{equation}
\label{radioloudness}
R=\left[\frac{L_{5\,{\rm GHz}}(\nu)}{L_{{\rm B}}(\nu)}\right],
\end{equation}
where $L_{5\,{\rm GHz}}(\nu)$ and $L_{\rm B}(\nu)$ are the rest-frame monochromatic luminosities at $5\,{\rm GHz}$ and in the optical B band, respectively.}
We converted the monochromatic flux at $1.4\,{\rm GHz}$, reported in the final catalog of the VLA-COSMOS Deep project (see for more detail Schinnerer et al. 2009, submitted, and \citealt{bondi08}) to $L_{5\,{\rm GHz}}(\nu)$ assuming $f(\nu)\propto \nu^{-\alpha}$ with $\alpha=0.7$. We excluded from the sample $51$ radio-loud Type 1 AGN (39 AGN from the spectroscopic sample and 12 sources with photometric redshift) with a value of radio-loudness $R \geqslant 10$.

The final Type 1 AGN sample used in our analysis, therefore, comprises $545$ X-ray selected AGN ($322$ from the spectro-z sample and $223$ from the photo-z sample) spanning a wide range of redshifts ($0.04<\z<4.25$) and X-ray luminosities ($40.6 \leq \Log \Lhard \leq 45.3$). The selection criteria are summarized in Table \ref{tbl-0}.

Assuming that the fraction of misclassificated sources in the spectro-z sample (17\%) and the fraction of missing Type 1 sources (10\%) could be applied to the photometric sample, we are able to estimate the uncertainties associated to the photo-z sample.
If this were the case, the incompleteness and the contamination on the total sample are very low, namely about 4\% and 7\%, respectively.

The redshift distributions of the total, spectroscopic and photometric samples are presented in Figure \ref{histz}. The median redshift of the total sample is 1.57 (the mean redshift is 1.66, with a dispersion of 0.75). The median redshift of the spectro-z sample is 1.54, while the median of the photo-z sample is 1.66.

\section{Rest-frame monochromatic fluxes and \\Spectral Energy Distributions}
\label{Rest-frame monochromatic fluxes and Spectral Energy Distributions}
To obtain rest-frame monochromatic luminosities at $2$ keV and $2500\textrm{\small\AA}$ and estimate bolometric luminosities, we used all the multi-color information as compiled by Brusa et al (2009). The catalog includes multi-wavelength data from mid infrared to hard X-rays: MIPS 24 $\mu$m GO3 data (Le Floc'h et al. 2009, ApJ submitted), IRAC flux densities (\citealt{sanders07}), near-infrared  CFHT/K-band data (\citealt{mccraken08}), HST/ACS F814W imaging of the COSMOS field (\citealt{koekemoer07}), optical multiband photometry (SDSS, Subaru, \citealt{capak07}) and near- and far-ultraviolet bands with GALEX (\citealt{2007ApJS..172..468Z}).
More specifically, the number of detections at $24\,\mu m$ is 472; for the 73 undetected sources, we consider 5$\sigma$ upper limits of $0.08 \, \rm m Jy$. Most of the sources are detected by IRAC: 545, 543, 544 and 543 at $3.6 \, \mu m$, $4.5 \, \mu m$, $5.8 \, \mu m$ and $8.0 \,\mu m$ band, respectively (all Type 1 AGN were detected in the 3.6$\mu m$ IRAC band). For the undetected sources we consider 5$\sigma$ upper limits of $1.7 \, \mu \rm Jy$, $11.3 \, \mu \rm Jy$ and $14.6 \, \mu \rm Jy$ at $4.5 \, \mu m$, $5.8 \, \mu m$ and $8.0 \,\mu m$, respectively (see Table 3 of \citealt{sanders07}). Only very small objects went undetected in the optical and near infrared bands: only 2 upper limits in the $\z^+$ band; 1 upper limit in the $B_{J}$, $V_{J}$ and $r^+$ bands; 8 upper limits in both $i^{*}$ and $u^{*}$ CFHT bands; 9 in the $K_{S}$ CFHT band and 31 in the $J$ UKIRT band. The observations are not simultaneous, as they span a time interval of about 5 years: 2001 (SDSS), 2004 (Subaru and CFHT) and 2006 (IRAC). In order to reduce variability effect, we selected the bands closest in time to the IRAC observations (i.e., we excluded SDSS data).
In Table \ref{tbl-2} we list the main X-ray and optical properties of the sample. The data for the SED computation were blueshifted to the rest-frame and no K-correction has been applied.
Galactic reddening has been taken into account: we used the selective attenuation of the stellar continuum $k(\lambda)$ taken from Table 11 of \citet{capak07}. Galactic extinction is estimated from \citet{schlegel98} for each object in the $\MC$ catalog. Count rates in the 0.5-2 keV and 2-10 keV are converted into monochromatic X-ray fluxes in the observed frame at 1 and 4 keV, respectively, using a Galactic column density $N_H = 2.5 \times 10^{20}\,cm^{-2}$ (see \citealt{cappelluti09}), and assuming a photon index $\Gamma_x=2$ and $\Gamma_x=1.7$, for the soft and hard band, respectively.

\begin{figure}
 \resizebox{\hsize}{!}{\includegraphics{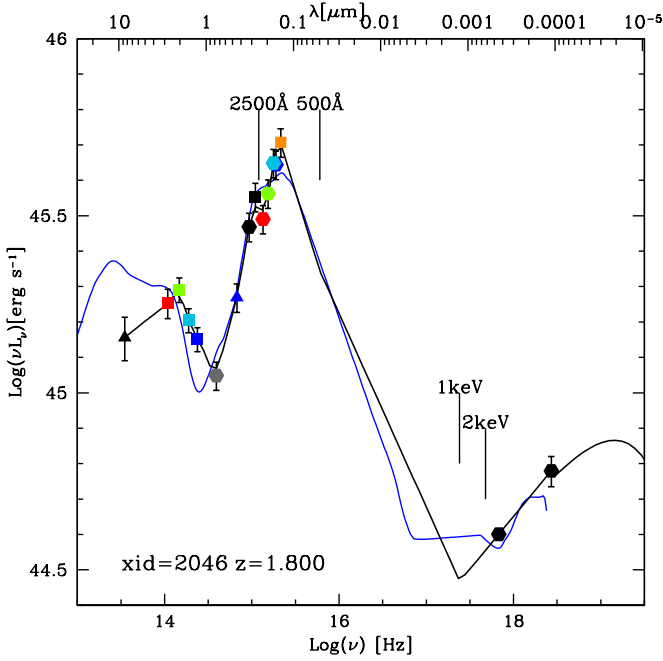}}
 \caption{The SED of a spectroscopically identified QSO at $z=1.8$ (XID=2046, \textit{black line}), compared with the mean SED of Elvis et al. (1994, \textit{blue line}). The rest-frame data, used to construct the SED of XID=2046 are, from left to right: black triangle, $24\, \mu m$ MIPS; red square, $8.0\, \mu m$ IRAC; green square, $5.8\, \mu m$ IRAC; cyan square, $4.5\, \mu m$ IRAC; blue square, $3.6\, \mu m$ IRAC; grey hexagon, K CFHT; blue triangle, J UKIRT; black, red, green, cyan and blue hexagons represent $\z^+$, $r^+$, $g^+$, $V_{J}$ and $B_{J}$ SUBARU bands, respectively; black and orange squares represent $i^{*}$ and $u^{*}$ CFHT bands, respectively; black hexagons soft X-ray and medium-hard X-ray luminosity.}
 \label{singlesed1}
\end{figure}
\begin{figure}
 \resizebox{\hsize}{!}{\includegraphics{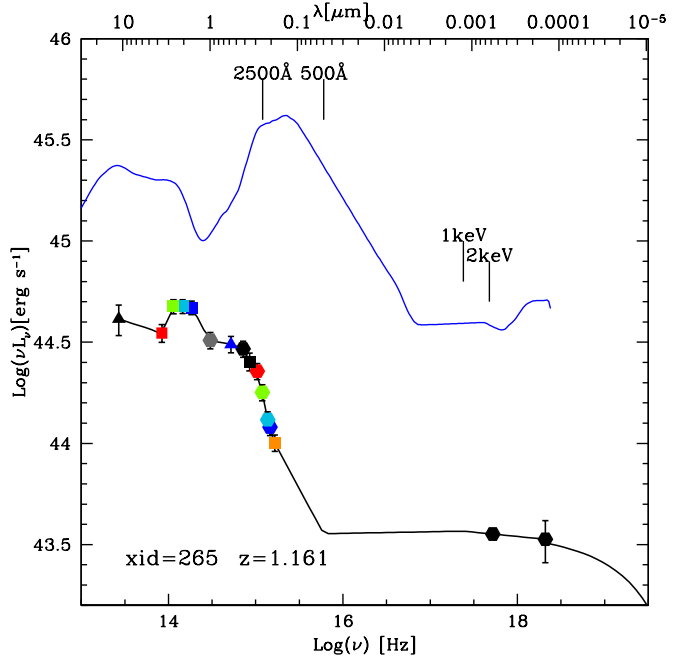}}
 \caption{The SED of a spectroscopically identified QSO at $z=1.161$ (XID=265, \textit{black line}). Keys are as in Fig. \ref{singlesed1}.}
 \label{singlesed2}
\end{figure}

\subsection{The SED Computation}
\label{The SED Computation}
With a procedure similar to that adopted by Elvis et al. (2009, in preparation), we assembled SEDs for the 545 Type 1 AGN. 
First of all, we plotted the information of all the sources, from $24\, \mu m$ to the hard X-ray band (the few upper limits in the optical and near-infrared are not considered), in the $\Log\nu-\Log[\nu L_\nu]$ rest-frame plane.
For each source we consider the rest-frame luminosity and corresponding frequency in each of the available bands. We determine a first order polynomial between two adjacent points, and the resulting function is then sampled with a step of $\Delta \Log \nu=0.085$. In this way a ``first order" SED, where every point is separated by 0.085 in log frequency, is created. The resulting SED is ``smoothed" using a fourth order polynomial interpolation and re-sampled with a step of $\Delta \Log \nu=0.01$. 
This choice is motivated by the fact that a single interpolation with a high-order polynomium could introduce spurious features in the final SED. 
This procedure allows us to build densely sampled SEDs at all frequencies, to extract accurate $\Alambda$ and 2 keV rest-frame monochromatic luminosities at all redshifts, to estimate bolometric luminosities and bolometric corrections.

For the computation of the bolometric luminosity we need to extrapolate the data in the UV to X-ray gap and at high X-ray energies. From the rest-frame UV luminosity data point at the highest frequency in each SED, we assume a power law spectrum to 500$\text{\AA}$, as measured by HST observations for radio-quiet AGN ($f_\nu\propto \nu^{-1.8}$, see \citealt{zheng97}). We then linearly connect the UV luminosity at 500 $\text{\AA}$ to the luminosity corresponding to the frequency of 1 keV. We extrapolate from the X-ray data point to the luminosity at 1 keV computing the slope, $\alpha_x=\Log(f_{4\,keV}/f_{1\,\rm keV})/\Log(\nu_{4\,keV}/\nu_{1\,keV})$ ($\alpha_x=\Gamma_x-1$). Finally, the X-ray spectrum is extrapolated at higher energies using the observed slope $\alpha_x$, and introducing an exponential cut-off at 200 keV, $f(E)\sim E^{-\alpha_x} e^{-E/200 \rm keV}$ (\citealt{2007A&A...463...79G}). Hence, the bolometric luminosity is computed integrating the extrapolated SED in the $\Log\nu-\Log[\nu L_\nu]$ rest-frame plane. Our bolometric luminosities are in agreement with those computed by Elvis et al. (2009). In Figure \ref{singlesed1} and \ref{singlesed2} we show two examples of SED used for the calculation of the bolometric luminosities and the rest-frame monochromatic luminosities of the Type 1 AGN sample. Given the selection criteria, the SEDs of the objects in our sample are quite different, ranging from blue QSO with an SED close to that reported by \citet{1994ApJS...95....1E} for radio-quiet AGN, to objects with red SED possibly due to intrinsic absorption and/or host-galaxy contamination. We will discuss in Sections \ref{Intrinsic Extinction} and \ref{Host Galaxy Contamination} how we take the possible effects of reddening and host-galaxy contamination into account in our analysis. A complete Type 1 AGN SED atlas will be presented in Elvis et al. (2009).

\begin{table*}[ct]
\caption{Optical and X-ray properties of the total sample. \label{tbl-2}}
\begin{tabular}{rcccccccrr}
\hline\hline\noalign{\smallskip}
 XID & Spectroscopic & Photometric   & $\Log \lo$ &  $\Log \lx$ & $\alphaox$  & $\Log \Lhard$ & $\Lbolmu$ & $k_{bol}$ & Class$^{\mathrm{a}}$   \\
     & redshift  &  redshift  & $\rm{[erg\,s^{-1}Hz^{-1}]}$ & $\rm{[erg\,s^{-1}Hz^{-1}]}$  & &$\rm{[erg\,s^{-1}]}$   & $[L_{\odot}]$ &  \\
\noalign{\smallskip}\hline\noalign{\smallskip}
   1    &  0.373  & 0.37$^{+0.03}_{-0.01}$  & 29.49   &   26.04  &  1.32    &  44.00  &  11.53  &  12.94  &    28 \\
   2    &  1.024  & 1.05$^{+0.03}_{-0.03}$  & 29.86   &   26.95  &  1.11    &  44.97  &  12.30  &   8.09  &    30 \\
   3    &  0.345  & 0.36$^{+0.02}_{-0.02}$  & 29.22   &   26.00  &  1.24    &  43.98  &  11.44  &  11.15  &    29 \\
   6    &  0.360  & 0.32$^{+0.02}_{-0.02}$  & 29.22   &   25.43  &  1.45    &  43.39  &  11.32  &  32.60  &    20 \\
   8    &  0.699  & 0.71$^{+0.01}_{-0.03}$  & 30.00   &   26.27  &  1.43    &  44.09  &  11.98  &  29.60  &    26 \\
  10    &  0.689  & 0.68$^{+0.02}_{-0.02}$  & 29.24   &   25.92  &  1.28    &  43.71  &  11.30  &  14.93  &    22 \\
  13    &  0.850  & 0.84$^{+0.02}_{-0.02}$  & 29.54   &   26.06  &  1.34    &  43.98  &  11.70  &  20.06  &    24 \\
  15    &  2.033  & 2.01$^{+0.03}_{-0.03}$  & 30.83   &   27.11  &  1.43    &  45.11  &  12.95  &  26.33  &    30 \\
  16    &  0.667  & 0.61$^{+0.03}_{-0.01}$  & 29.32   &   25.82  &  1.34    &  43.78  &  11.38  &  15.14  &    19 \\
  17    &  1.236  & 1.28$^{+0.02}_{-0.02}$  & 30.77   &   26.77  &  1.53    &  44.70  &  12.75  &  43.09  &    29 \\
\hline\noalign{\smallskip}                      
\end{tabular}                                   
                                                
\begin{list}{}{Notes---This table is presented entirely in the electronic edition; a portion is shown here for guidance.}
\item[$^{\mathrm{a}}$]The SED model of the best-fit template is coded from 1 to 30, as detailed in S09.
\end{list}                                      
\end{table*}                                    
                                                

\section{Statistical Analysis}                  
\label{Statistical Analysis}                    
To study the possible correlations between $\alphaox$, $\lo$, $\lx$ and redshift, we used the Astronomy Survival Analysis software package (ASURV rev. 1.2; \citealt{isobe90}; \citealt{lavalley92}). ASURV implements the bivariate data-analysis methods and also properly treats censored data using the survival analysis methods (\citealt{feigelsonnelson85}; \citealt{isobe86}). We have employed both full parametric estimate and maximized (EM) regression algorithm and semiparametric Buckley-James regression algorithm (Buckley \& James 1979) to perform the linear regression of the data.  The EM regression algorithm is based on the ordinary least-squares regression of the dependent variable Y against the independent variable X (OLS[Y$|$X]). The regression line is defined in such a way that it minimizes the sum of the squares of the Y residuals. Since in our analysis the choice of the independent variable is not straightforward, we have also used the inverse of OLS(Y$|$X) (OLS[X$|$Y]) and we then calculated the bisector of the two regression lines (see \citealt{isobe90}). We only present the findings from the EM regression algorithm, since in all cases the results from the Buckley-James regression algorithm agreed within the errors; we also report the findings from OLS bisector. Moreover, we used the partial-correlation analysis method to compute the correlation between two variables, checking the effect of an additional parameter which the two variables depend on. This method is implemented in the FORTRAN program PARTIAL\_TAU, available from the Penn State Center of Astrostatistics, and uses the methodology presented in \citet{akritas96}. The procedure is based on Kendall $\tau$-statistic that properly handles censored data.
                                                
We summarize the results for partial-correlation analysis in Table \ref{tbl-3}.
\begin{table}[ct]                               
\caption{Correlations and their significance from Kendall-$\tau$ statistics (K-$\tau$) and from Partial Kendall-$\tau$. \label{tbl-3}}
\begin{tabular}{lccc}                           
\hline\hline\noalign{\smallskip}                
Correlation  & K-$\tau$    &  Controlling variable   &   Partial K-$\tau$\\
 & ($\sigma$) &  & ($\sigma$) \\                
\hline\noalign{\smallskip}
  \multicolumn{4}{c}{Total} \\
\hline\noalign{\smallskip}                      
   $\alphaox-\lo$ & 17  & $\z$   & 21 \\        
   $\alphaox-\lx$ & 1.2  & $\z$  & 1.5 \\    
   $\alphaox-\z$   & 7  & $\lo$ & $<0.1$ \\   
\hline\noalign{\smallskip}
  \multicolumn{4}{c}{Spectro-z} \\
\hline\noalign{\smallskip}                      
   $\alphaox-\lo$  & 12  & $\z$  & 14 \\        
   $\alphaox-\lx$  & 0.2  & $\z$  & 2.3 \\    
   $\alphaox-\z$   & 5  & $\lo$ & $<0.1$ \\   
\hline\noalign{\smallskip}
  \multicolumn{4}{c}{Photo-z} \\
\hline\noalign{\smallskip}                      
   $\alphaox-\lo$  & 10  & $\z$  & 12 \\        
   $\alphaox-\lx$  & 0.2  & $\z$  & 3 \\    
   $\alphaox-\z$   & 5  & $\lo$ & 0.7 \\   
\hline\noalign{\smallskip}
                      
\end{tabular}                                   
\end{table}                                     
                                                
\begin{figure*}                                 
\centering                                      
 \includegraphics[width=13.2cm,height=13.2cm]{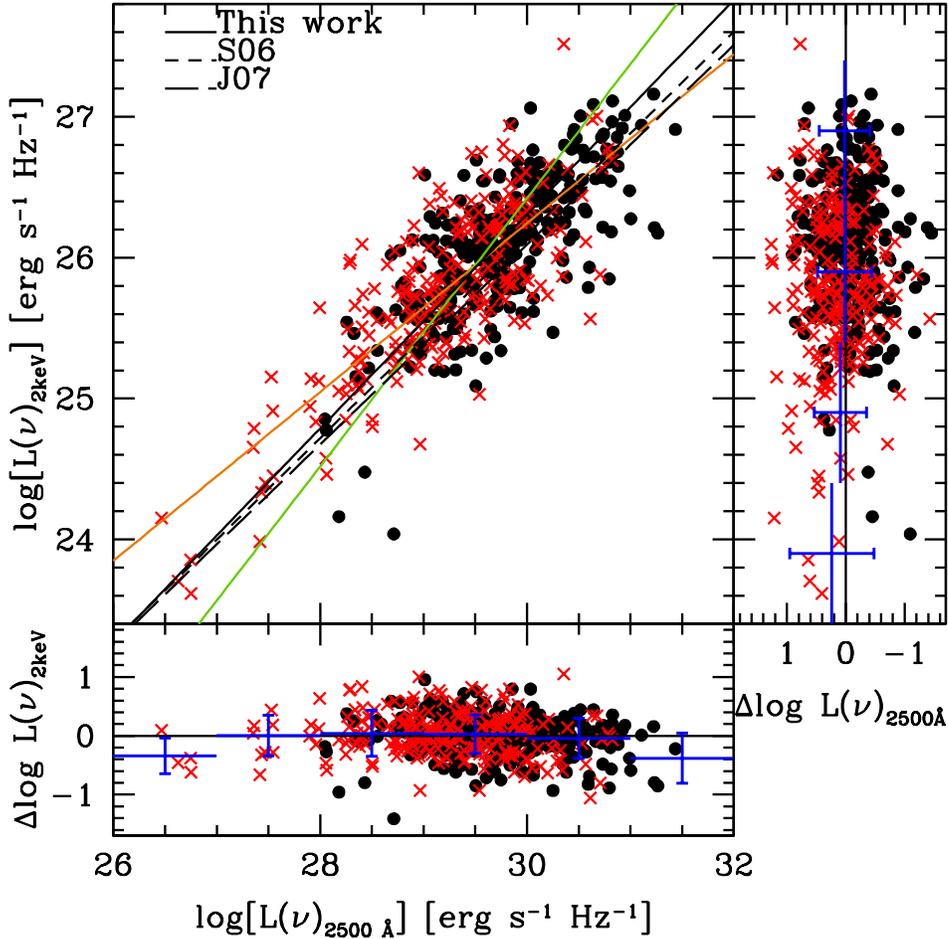}
 \caption{Values of the rest-frame monochromatic luminosity $\lx$ plotted versus the rest-frame $\lo$ monochromatic luminosity for our X-ray selected sample: $322$ spectroscopic sources (\textit{circles}) and $223$ sources with photometric redshift (\textit{crosses}). The solid black line shows the best-fit relation that we found using the OLS bisector algorithm (see eq. [\ref{lo_lx_bisector}]). For comparison, the best-fit derived by S06 (\textit{short-dashed line}) and the best-fit from J07 (\textit{long-dashed line}) are also shown. The orange and the green line represent equations (\ref{lo_lx_loind}) and (\ref{lo_lx_lodip}), respectively. The lower panel and the panel on the right show the residuals of log $\lx$ and log $\lo$ with respect to the best-fit lines given by eq. [\ref{lo_lx_loind}] and [\ref{lo_lx_lodip}]), respectively. The error bars represent the mean and the $1\sigma$ dispersion of the residuals for each $\Delta \Log(\lo)=1$ bin.}
 \label{lxlorq}                                 
\end{figure*}                                   
                                                
\subsection{$\lo$ vs $\lx$}                     
\label{lolxsection}
Previous studies on optically selected AGN reported a relationship between $\lx$ and $\lo$ in the form $\lx \propto \lo^\beta$, with the best-fit values of the exponent between $0.7 \div 0.8$ (e.g. \citealt{avnitananbaum82}, 1986; \citealt{chanan83}; \citealt{krisscanizares85}; \citealt{andersonmargon87}; \citealt{wilkes94}; \citealt{yuansiebertbrink98};\citealt{vignali03}; \citealt{strateva05}; S06; J07; see also \citet{lafranca95}, who found $\beta\sim1$). For the best-fit parameters using OLS(Y$|$X) (i.e. treating $\lo$ as the independent variable) we find
\begin{equation}                                
\label{lo_lx_loind}                             
\Log \lx=(0.599\pm0.027) \Log \lo + (8.275\pm0.801)
\end{equation}
while treating $\lo$ as the dependent variable (i.e., OLS(X$|$Y)) and inverting the resulting best-fit, we find
\begin{equation}
\label{lo_lx_lodip}
\Log \lx=(0.952\pm0.033) \Log \lo - (2.138\pm0.975).
\end{equation}
We then compute the bisector of the two regression lines as described by \cite{isobe90} and find
\begin{equation}
\label{lo_lx_bisector}
\Log \lx=(0.760\pm0.022) \Log \lo + (3.508\pm0.641)
\end{equation}
with a dispersion of 0.37.
The difference of the best-fit $\beta$ with respect to a linear correlation (i.e., $\beta=1$) is highly significant ($11\,\sigma$), this result confirms the non linear correlation between $\lx-\lo$.
This implies not only that $\alphaox$ must be dependent on optical luminosity, but also that optically bright AGN emit less X-ray (per unit UV luminosity) than optically faint AGN.
Comparing our results with those obtained from optically selected samples, we find that our slope is consistent within $2\,\sigma$ with those of S06 sample ($\beta_{S06}=0.721\pm0.011$) and J07 sample ($\beta_{J07}=0.709\pm0.010$), while it is significantly smaller than the slope of $1.117\pm0.017$ found by \citet{green09}. Moreover, our slope is similar to the value found by \citet{2009arXiv0909.0464S} who used an X-ray selected sample composed by 267 broad-line AGN ($\beta=0.870$).
\rev{Treating separately the spectro-z and photo-z samples, we found that the best-fit slope for the bisector is $0.782\pm0.033$ with a normalization $2.815\pm0.989$ and $0.786\pm0.033$ with a normalization $2.824\pm0.958$, respectively. The spectro-z sample is limited to relatively bright $2500 \,\text{\AA}$ luminosities, while the photo-z sample extends the luminosity range of about 1.5 dex towards lower $\lo$ values. It is worth noting that there is a significant difference between the normalization in eq. (\ref{lo_lx_bisector}) and the normalizations computed treating separately the spectro-z and the photo-z samples. The percentual difference between the predicted $\lx$ using the best-fit relation for the total sample and the best-fit for the spectro-z sample is $\sim 10\%$.
Moreover, considering only the results from the spectro-z sample, we would slightly underestimate the dispersion in the bisector.
These considerations justified the choice of adding the photo-z sample in order to reduce possible biases associated to the spectro-z sample.}
In Figure \ref{lxlorq} we show monochromatic X-ray luminosities as a function of monochromatic UV luminosities for our X-ray selected sample. For comparison, we plot the $\lx-\lo$ relations from S06 and J07. In the bottom panel, we report residuals from equation (\ref{lo_lx_loind}), while on the right panel, residuals from eq. (\ref{lo_lx_lodip}) with the mean and the $1\sigma$ dispersion computed for each bin with $\Delta\Log L=1$ are reported.

\subsection{$\alphaox$ vs $\lo$}
\label{alphaoxlo}
\begin{figure}
 \resizebox{\hsize}{!}{\includegraphics{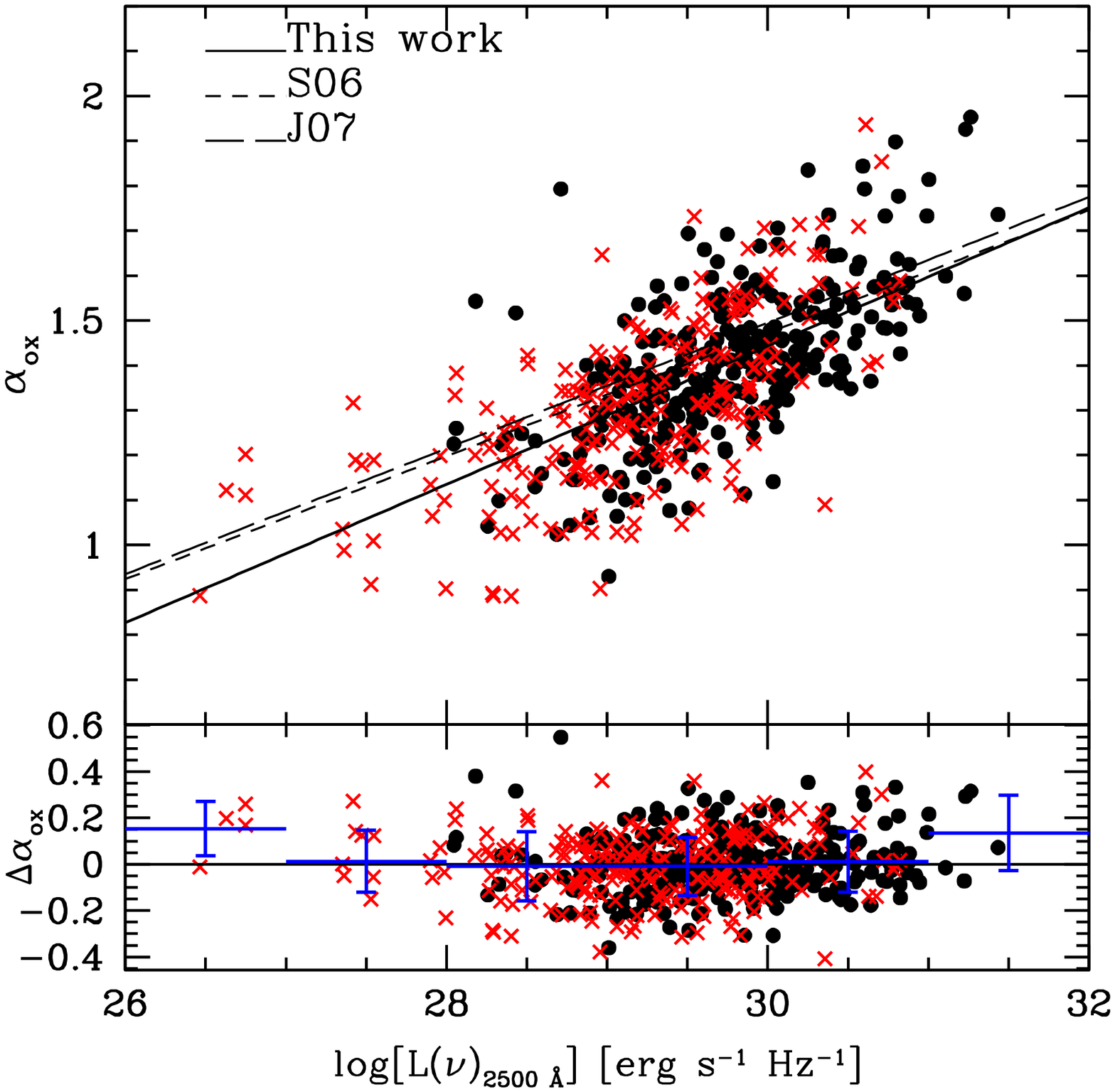}}
 \caption{Plot of the $\alphaox$ vs. the rest-frame monochromatic luminosity $\lo$ for our X-ray selected sample: $322$ spectroscopic sources (\textit{circles}) and $223$ sources with photometric redshift (\textit{crosses}). The solid line represents our best-fit given by equation (\ref{aoxlo_loind}). For comparison, the best-fit derived by S06 (\textit{short-dashed line}) and from J07 (\textit{long-dashed line}) are shown. The lower panel shows the residuals ($\Delta \alphaox$) of our best-fit relation. The error bars represent the mean and the $1\sigma$ standard deviation of the mean of the residuals for each $\Delta \Log(\lo)=1$ bin.}
 \label{loaoxrq}
\end{figure}

Previous X-ray studies of AGN using optically selected samples (e.g.; \citealt{vignali03}, \citealt{strateva05}, S06, J07) revealed a highly significant correlation between $\alphaox$ and the $2500 \,\text{\AA}$ monochromatic luminosity, with a slope of $\sim0.14$. We apply EM regression to our X-ray selected Type 1 sample and confirm the $\alphaox-\lo$ correlation at the $17 \,\sigma$ significance level. The correlation becomes stronger when taking into account the effect of redshift using Kendall-$\tau$ partial-correlation analysis ($\sim21 \,\sigma$). The best-fit relation for $\alphaox-\lo$, using OLS(Y$|$X) (i.e. treating $\lo$ as the independent variable), is
\begin{equation}
\label{aoxlo_loind}
 \alphaox(\lo)=(0.154\pm 0.010) \Log \lo - (3.176\pm0.223),
\end{equation}
with a dispersion of 0.18.
\rev{For the spectro-z sample, we found a slope of $0.166\pm0.011$ with a normalization of $-3.541\pm0.335$, while for the photo-z sample the best-fit slope is $0.142\pm0.012$ with a normalization $-2.831\pm0.345$.}
In Figure \ref{loaoxrq} we plot $\alphaox$ versus $\lo$ for our sample. For comparison, we plot the best-fit linear regression found in S06 (\textit{short-dashed line}) and J07 (\textit{long-dashed line}). Our slope is statistically consistent within $\sim 1.5 \sigma$ with the slopes of the S06 and J07 best-fits. Moreover, our slope is consistent, within $1.6\sigma$ and $1\sigma$, with those published recently by \citet{kelly08} ($0.12\pm0.02$) and \citet{2009ApJS..183...17Y} ($0.153\pm0.012$), respectively. Differently, \citet{green09} and \citet{2009arXiv0909.0464S} found a flatter slope of $0.061\pm0.009$ and $0.065\pm0.019$, respectively. The residual $\Delta \alphaox$, defined as
\begin{equation}
 \Delta \alphaox = \alphaox - \alphaox(\lo),
\end{equation}
where $\alphaox$ values are measured from the observations, are shown in the bottom panel of Fig. \ref{loaoxrq}. The bars represent the mean and the $1\,\sigma$ dispersion for $\Delta \Log(\lo)=1$ bins.

\begin{figure}
 \resizebox{\hsize}{!}{\includegraphics{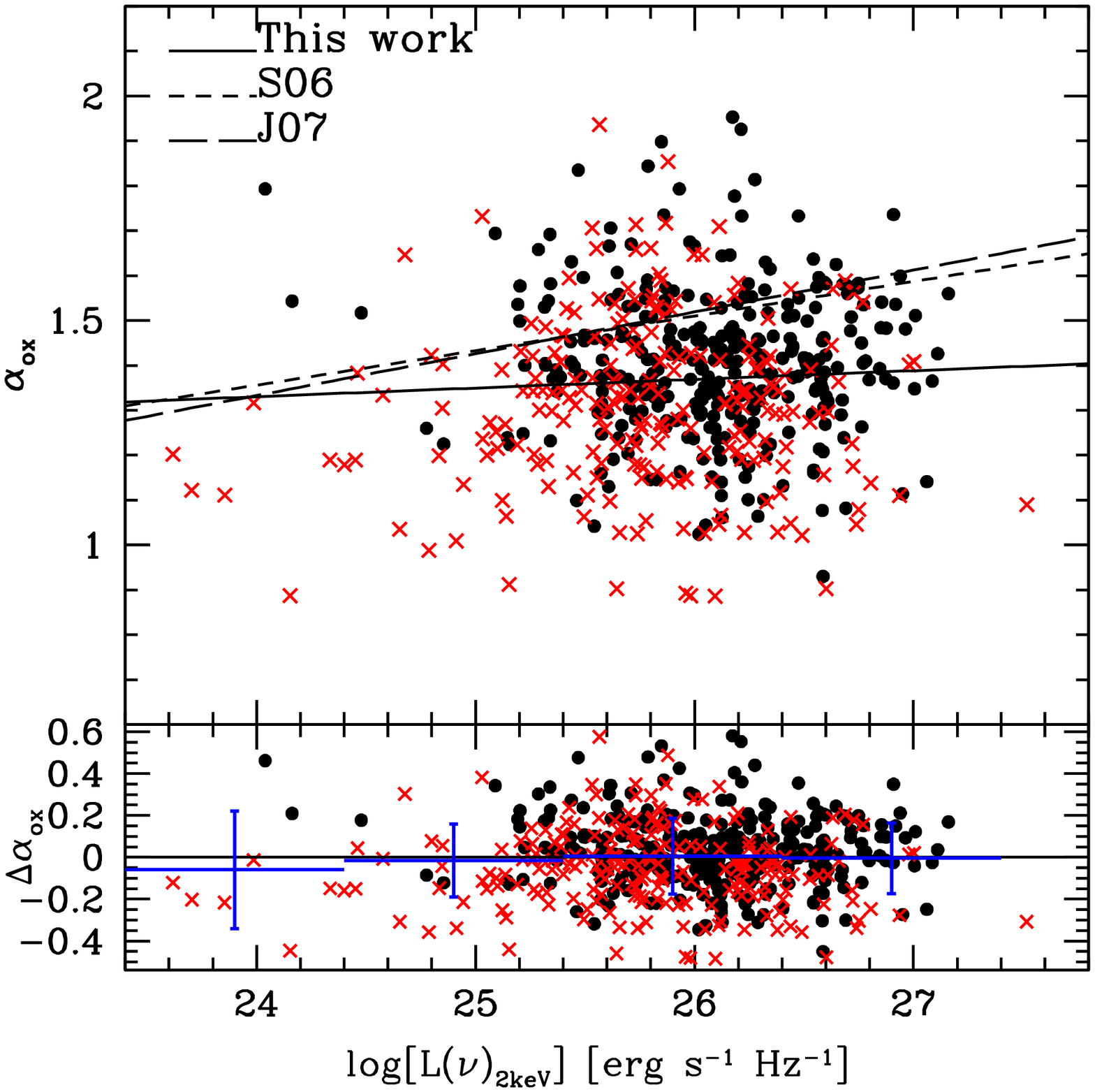}}
 \caption{Plot of $\alphaox$ vs. the rest-frame monochromatic luminosity $\lx$ for our X-ray selected sample: $322$ spectroscopic sources (\textit{circles}) and $223$ sources with photometric redshift (\textit{crosses}). The solid line represents our best-fit given by equation (\ref{aoxlx_lxind}). For comparison, the best-fit derived by S06 (\textit{short-dashed line}) and from J07 (\textit{long-dashed line}) are also shown. The lower panel shows residuals ($\Delta \alphaox$) from our best-fit relation, while error bars represent the mean and the $1\sigma$ dispersion of the residuals for each $\Delta [\Log \lx]=1$ bin.}
 \label{lxaoxrq}
\end{figure}

\begin{figure}
 \resizebox{\hsize}{!}{\includegraphics{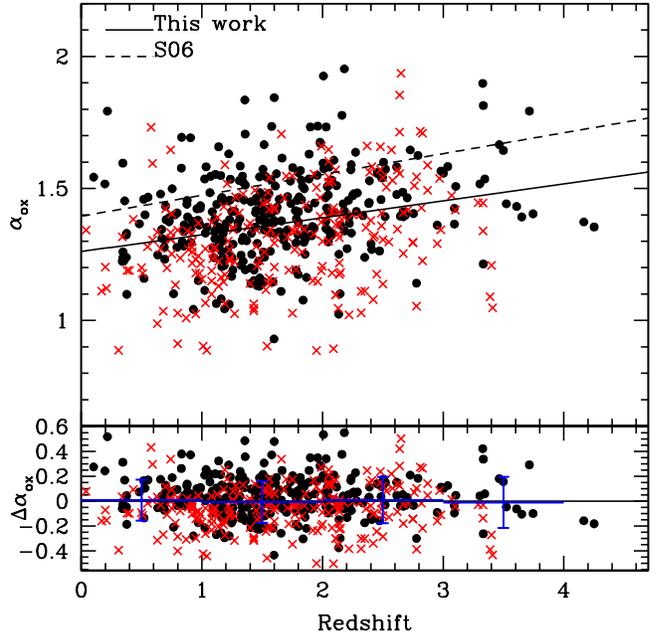}}
 \caption{Plot of $\alphaox$ vs. redshift for our X-ray selected sample: $322$ spectroscopic sources (\textit{circles}) and $223$ sources with photometric redshift (\textit{crosses}). The solid line represents our best-fit given by equation (\ref{aoxz_zind}). For comparison, we show also the best-fit derived by S06 (\textit{short-dashed line}). The lower panel shows residuals ($\Delta \alphaox$) from our best-fit relation, while error bars represent the mean and the $1\sigma$ dispersion of the residuals for each $\Delta \z=1$ bin.}
 \label{zaoxrq}
\end{figure}

\subsection{$\alphaox$ vs $\lx$ and vs redshift}
\label{alphaoxlxredshift}
We find no significant correlation between $\alphaox$ and $\lx$ (e.g.; \citealt{krisscanizares85}; \citealt{avnitananbaum86}; \citealt{wilkes94} and \citealt{yuansiebertbrink98}). For comparison with S06 and J07, we perform the EM regression method and the best-fit parameters for the $\alphaox-\lx$ relation (i.e. treating $\lx$ as the independent variable) are 
\begin{equation}
\label{aoxlx_lxind}
 \alphaox(\lx)=(0.019\pm0.013) \Log \lx + (0.863\pm0.344).
\end{equation}
\rev{For the spectro-z sample the slope is $0.012\pm0.020$, while for the photo-z sample it is $-0.020\pm0.019$.}
Analyzing in the same way the $\alphaox-\z$ relation (i.e. treating $\z$ as the independent variable), we find
\begin{equation}
\label{aoxz_zind}
 \alphaox(\z)=(0.064\pm0.010)\; \z + (1.261\pm0.018).
\end{equation}
The significance of the correlation between $\alphaox$ and redshift is $7\,\sigma$. This may suggest a not negligible evolution of $\alphaox$ with redshift, but if we account for the effect of the optical luminosity using partial Kendall-$\tau$ statistics, the significance of the correlation disappears ($<0.1\,\sigma$). In Figure \ref{lxaoxrq} and \ref{zaoxrq} we show $\alphaox$ as a function of $\lx$ and redshift, respectively, where the solid line represents the best-fit relation (eq. [\ref{aoxlx_lxind}] and [\ref{aoxz_zind}]). In the bottom panel of each plot we show the residuals and the $1\sigma$ dispersion in bins of $\Delta \z=\Delta \Log \lx=1$.
The absence of a correlation between $\alphaox$ and $\lx$ is consistent with the fact that the correlation between $\lo$ and $\lx$, treating $\lx$ as the independent variable (eq.[\ref{lo_lx_lodip}]), is very close to be linear (i.e., $\beta\sim 1$).
From the partial correlation analysis we showed that there is no correlation between $\alphaox$ and redshift, once the effect of optical luminosity is properly removed.
The energy mechanisms that generate the broad-band emission in AGN do not vary, in an appreciable way, over cosmic time. This is in agreement with most of previous works (but see \citet{yuansiebertbrink98} and \citet{bechtold03} who claimed that the primary correlation comes out upon redshift).
The difference in the normalization between our $\alphaox-z$ best-fit relation and S06 best-fit relation is probably due to the different optical luminosity range spanned by our data and S06 data. The average optical luminosity of the S06 sample is more than 1 decade higher than ours. Because of the correlation between $\alphaox$ and $\lo$, this implies a stepeer $\alphaox$ at any given redshift.
\rev{For the spectro-z sample we found that the slope is $0.058\pm0.012$, while for the photo-z sample it is $0.079\pm0.016$.}

\begin{figure}
 \resizebox{\hsize}{!}{\includegraphics{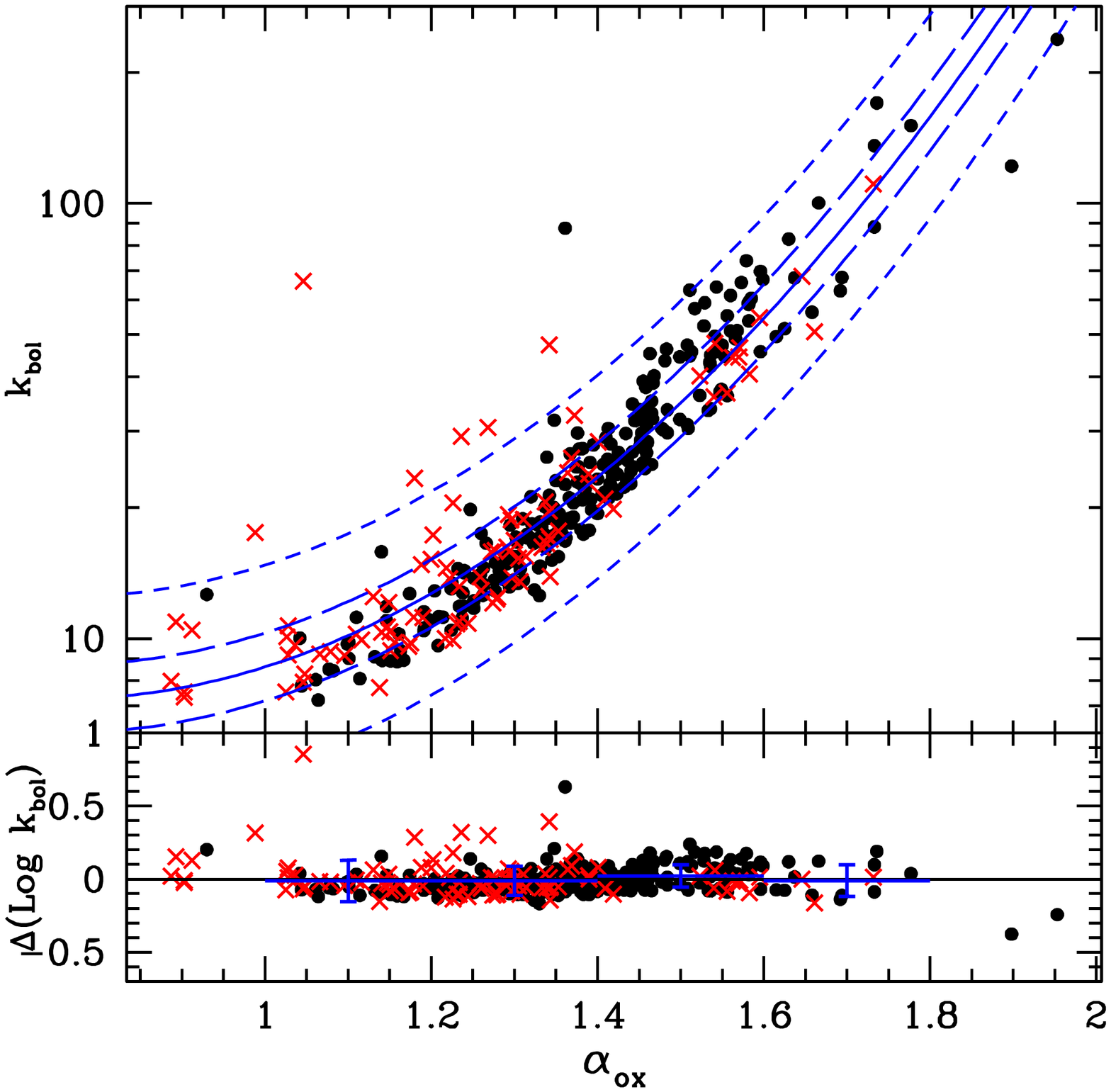}}
 \caption{Plot of the hard X-ray bolometric corrections against $\alphaox$ for the 343 Type 1 AGN with detection in both soft and hard bands. The blue solid line represents the best-fit relation from eq.(\ref{aoxkbol343}), while the long and the short dashed lines represent the $1\sigma$ and the $3\sigma$ dispersion in the distribution of bolometric correction, respectively. The bottom panel shows residuals ($\Delta [\Log k_{bol,\,1\mu m}]$) from the best-fit relation, while error bars represent the mean and the $1\sigma$ dispersion of the residuals for each $\Delta (\Log k_{bol,\,1\mu m})=0.2$ bin.}
 \label{kbolaox}
\end{figure}

\subsection{Bolometric Corrections vs Eddington ratio}

In Figure \ref{kbolaox} we plot $\alphaox$ vs. $\kbol$. In order to better estimate the bolometric correction values we select a subsample of 343 AGN with detections in both soft and hard bands.
The best-fit relation is computed using a second-degree polynomial:
\begin{equation}
\label{aoxkbol343}
  \Log \kbol=1.561-1.853 \alphaox + 1.226 \alphaox^2.
\end{equation}
We quantify the $1\sigma$ ($\sim0.078$) and the $3\sigma$ ($\sim0.234$) dispersion in the $\Log \kbol-\alphaox$ relation using a $3.5\sigma$ clipping method.
\rev{A linear correlation is not a good description of the observed data points. The addition of a quadratic term significantly improves the fit quality ($\Delta \chi^2\sim 140$, significant at $\gtrsim 8\sigma$ confidence level according to an F-test).
A visual inspection of the individual SED of the sources which mostly deviate from a linear fit (i.e. those with $\alphaox\leqslant1.2$) indicate that their SEDs are not significantly different from those of the rest of the sample.}
We also note that the dispersion around the best-fit is very small
For 150 Type 1 AGN in our sample in the redshift range $0.196 \leq \z \leq 4.251$, an estimate of the BH mass is available from virial estimators (Peterson et al. 2004) using the \ion{Mg}{ii} line width (63 sources from Merloni et al. submitted; 63 sources from \citealt{trump09}); the \ion{H}{$\beta$} line width (16 sources from \citealt{trump09}) and the \ion{C}{iv} line width (8 sources from \citealt{trump09}).

The bolometric luminosity which enters in the calculation of the Eddington ratio, $\lambdaEdd$, is computed by integrating the individual SEDs from $1\mu m$ to 200 keV (see Section \ref{The SED Computation}). By neglecting the IR bump we avoid counting twice the UV emission reprocessed by dust. The hard X-ray bolometric correction, $\kbol$, is computed as the ratio between $\Lbolmu$ and the X-ray luminosity, $\Lhard$.

A remarkably good correlation is found between $\kbol$ and the Eddington ratio. Since the choice of the independent or dependent variable is not straightforward, we here computed the OLS bisector for the $\kbol-\lambdaEdd$ relation as already done for the $\lo-\lx$ relation.
We found that the best-fit parameters for the $\kbol-\lambdaEdd$ relation using OLS(Y$|$X) (i.e. treating $\lambdaEdd$ as the independent variable) are
\begin{equation}
\label{kbolledd_oyx}
 \Log \kbol(\lambdaEdd) = (0.273\pm0.045)\Log \lambdaEdd + (1.656\pm0.056),
\end{equation}
while the relation using OLS(X$|$Y) is
\begin{equation}
\label{kbolledd_oxy}
 \Log \kbol(\lambdaEdd) = (1.200\pm0.170)\Log \lambdaEdd + (2.600\pm0.180).
\end{equation}
We then compute the bisector of the two regression lines and we find
\begin{equation}
\label{kbolledd_bis}
 \Log \kbol(\lambdaEdd) = (0.643\pm0.043)\Log \lambdaEdd + (2.032\pm0.053).
\end{equation}
The absence of a correlation is excluded at the $15\sigma$ level.

\begin{figure}
 \resizebox{\hsize}{!}{\includegraphics{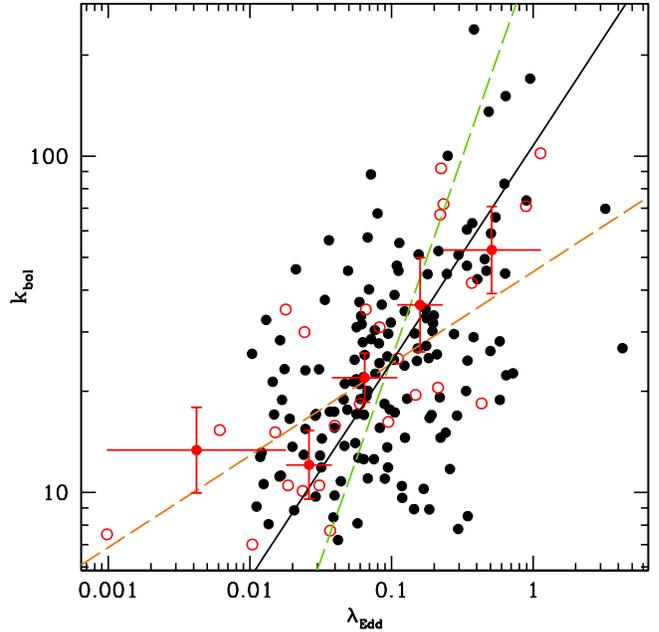}}
 \caption{Hard X-ray bolometric correction versus Eddington ratio for the 150 Type 1 AGN with BH mass estimate. The solid black line shows the best-fit relation that we found using the OLS bisector algorithm (see eq. [\ref{kbolledd_bis}]), while the orange and the green dashed lines represent equations (\ref{kbolledd_oyx}) and (\ref{kbolledd_oxy}), respectively. Red open circles represent the sample by VF09 (25 sources, see their Fig. 6 for details) and the corresponding bins with error bars.}
 \label{kboleddcheck}
\end{figure}

Figure \ref{kboleddcheck} shows $\kbol$ as a function of $\lambdaEdd$ for our sub-sample of Type 1 AGN and for the sample of local Seyferts by VF09, where radio-loud objects and low X-ray flux observations are removed. Points with error bars are obtained averaging the values of bolometric correction in each bin of $\lambdaEdd$, and standard errors at $1\sigma$ are plotted for comparison (see Section 4.3 of VF09 for details). The VF09 sample contains AGN with simultaneous optical, UV and X-ray data retrieved from the XMM-\textit{Newton} EPIC-pn and Optical Monitor (OM) archives, while the virial $\MBH$ is estimated using the reverberation mapping method (\citealt{peterson04}). 
It is worth noting that, although the methods to construct the SEDs and measure $\MBH$ are completely different from those adopted by VF09, the trend of increasing bolometric correction with Eddington ratio is confirmed, with mean $\langle\kbol\rangle\sim 22$ for $\lambdaEdd \leq 0.1$, $\langle\kbol\rangle\sim 27$ for $0.1 < \lambdaEdd \leq 0.2$ and $\langle\kbol\rangle\sim 53$ for $\lambdaEdd > 0.2$.

Equation (\ref{kbolledd_oyx}) is in agreement, within the errors, with the VF09 results, which are shown in Figure \ref{kboleddcheck}. The equation (\ref{kbolledd_oyx}), represented by the orange dashed line, and red bins are directly comparable, because of the treatment of the independent variable.

\begin{figure}
 \resizebox{\hsize}{!}{\includegraphics{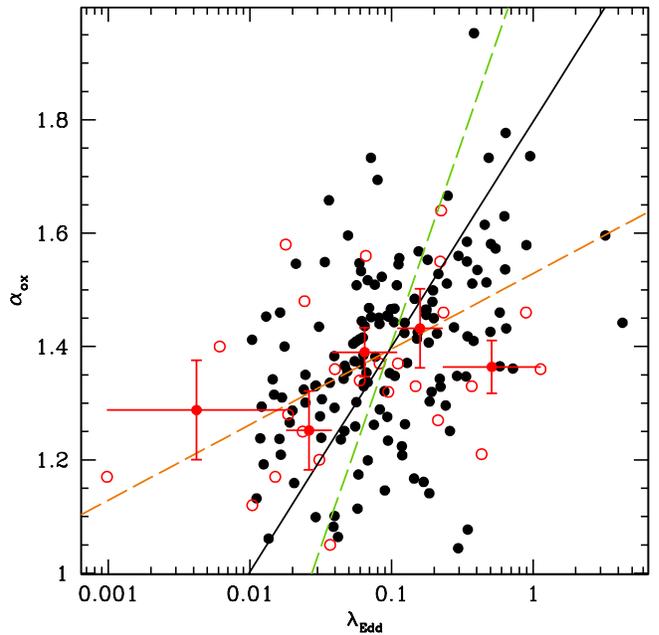}}
 \caption{Plot of $\alphaox$ as a function of Eddington ratio for the 150 Type 1 AGN and the corresponding bins with error bars. Key as in Figure \ref{kboleddcheck}. The solid black line shows the best-fit relation that we found using the OLS bisector algorithm (see eq. [\ref{aoxEddratio}]), while the orange and the green dashed lines represent equations (\ref{aox_ledd_oyx}) and (\ref{aox_ledd_oxy}), respectively.}
 \label{leddaox}
\end{figure}

Differently from VF09, we found a correlation also between $\alphaox$ and the Eddington ratio. In the same interval of $\lambdaEdd$ we have a larger number of sources, hence the presence of the correlation could be simply due to the better statistic, although the dispersion is rather large. In Fig. \ref{leddaox} we present $\alphaox$ against $\lambdaEdd$, where the best-fit relation using OLS(Y$|$X) is
\begin{equation}
\label{aox_ledd_oyx}
 \alphaox(\lambdaEdd)=(0.133\pm0.023) \Log \lambdaEdd + (1.529\pm0.028),
\end{equation}
while the best-fit relation using OLS(X$|$Y) is
\begin{equation}
\label{aox_ledd_oxy}
 \alphaox(\lambdaEdd)=(0.719\pm0.127) \Log \lambdaEdd + (2.124\pm0.132).
\end{equation}
Finally, the bisector of the two regression lines is
\begin{equation}
\label{aoxEddratio}
 \alphaox(\lambdaEdd)=(0.397\pm0.043) \Log \lambdaEdd + (1.797\pm0.047).
\end{equation}
The absence of a correlation is excluded at $\sim 9 \sigma$ level.

\section{Effects of reddening and host-galaxy light}
\label{Effects of reddening and host galaxy light}

As mentioned in Section \ref{The SED Computation} and discussed at length in Elvis et al. (2009, in prep.), the objects used in this analysis show a large variety of SEDs.
In addiction to objects with a ``typical" Type 1 AGN SED (see Fig. \ref{singlesed1}), there is also a not negligible number of objects which show a red optical-UV SED (see Fig. \ref{singlesed2}).
While some of them may be intrinsically red AGN, it is likely that for most of them this red optical-UV SED is due either to intrinsic absorption or to a significant contribution of emission from the host-galaxy, or both (see \citealt{richards03}, hereafter R03).
In the analysis presented in the previous Sections, we used the ``observed" $\lo$. If intrinsic absorption or contribution from the host-galaxy are not negligible, the used $\lo$ would be biased. In particular, the intrinsic $\lo$ would be higher in presence of reddening, while it would be smaller in presence of a significant contribution from the host-galaxy.
In the following we try to estimate the size of the effect of extinction and/or host-galaxy contribution to the relations between $\lo-\lx$ and $\alphaox-\lo$ discussed in Section \ref{lolxsection} and \ref{alphaoxlo}.

\subsection{Intrinsic Extinction}
\label{Intrinsic Extinction}    
\begin{figure}                  
 \resizebox{\hsize}{!}{\includegraphics{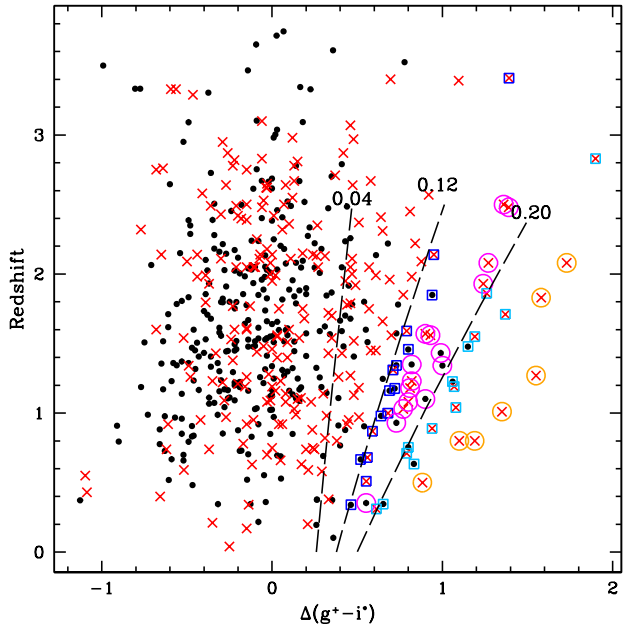}}
 \caption{Distribution of redshifts versus the relative $\Delta(g^+-i^*)$ color for the sample with $\z<4$. The dashed lines represent, from left to right, the expected loci for E(B-V)=0.04, 0.12, 0.20, respectively. Open circles and squares mark \textit{reddened} AGN: blue corresponds to E(B-V)$\sim$0.12 (16 AGN), magenta to E(B-V)$\sim$0.16 (16 AGN), cyan to E(B-V)$\sim$0.20 (14 AGN) and yellow to E(B-V)$\sim$0.24 (7 AGN).}
 \label{deltagiredshift}        
\end{figure}          
\begin{figure}        
 \resizebox{\hsize}{!}{\includegraphics{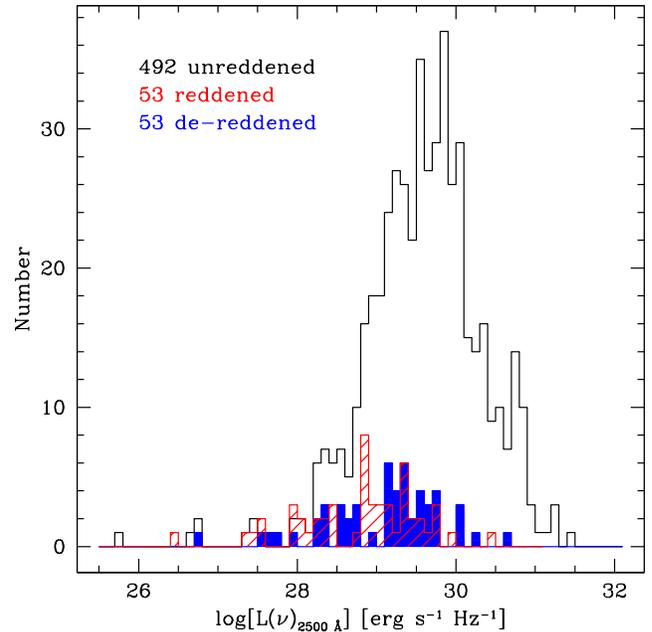}}
 \caption{Distribution of $\lo$ for the unreddened sample (\textit{open histogram}) and for the 53 sources with E(B-V)$\geq 0.12$, before (\textit{red hatched histogram}) and after reddening correction (\textit{blue filled histogram}).}
\label{lohistred}
\end{figure}          
Following the procedure described in R03, we estimate the underlying continuum color by subtracting the median colors of AGN in redshift bins of $\Delta \z = 0.2$ from the observed color of each source in that bin. We refer to $\Delta (g^+-i^*)=(g^+-i^*)-\langle g^+-i^*\rangle$ as the \textit{relative color} (see \citealt{richards01}), where the average $\langle g^+-i^* \rangle$ is computed in each redshift bin. All sources have been detected in the Subaru $g^+$ band, only 8 sources do not have $i^*$ CFHT magnitude. For these AGN we consider the detection in the $i^+$ Subaru band. In Figure \ref{deltagiredshift} we plot the redshift of the sources as a function of their relative color (see Fig. 6 in R03).
The $\Delta(g^+-i^*)$ distribution shows a large scatter with respect to the SDSS sample in R03. Given the quality of the COSMOS photometric data, this is not due to photometric errors, but likely to the different selection criteria: SDSS quasar candidates are selected using optical color-color selection, so the scatter in $\Delta(g^+-i^*)$ is smaller than in our X-ray selected sample. However, we can still use the same plot in order to identify possibly reddened sources. Under the assumption that all Type 1 AGN have the same continuum shape, the dashed lines show the expected change in relative color as a function of redshift for an SMC reddening law (\citealt{prevot84}) with E(B-V) = 0.04, 0.12 and 0.20 moving from left to right in the $\Delta(g^+-i^*)$ axis. Following R03, we define dust-reddened Type 1 AGN all the sources that lie to the right of the dashed line at E(B-V)=0.12. Using this definition, about 10\% of the total sample is affected by intrinsic absorption. For about 80\% of the sample, reddening is negligible, with E(B-V) less than 0.04. The 53 reddened Type 1 AGN have been divided in 4 subsamples which correspond to an average E(B-V) value of about 0.12, 0.16, 0.20 and 0.24 (see caption in Fig. \ref{deltagiredshift}). Monochromatic luminosities at $\Alambda$ of these objects were corrected using the SMC reddening law and the corresponding average value of E(B-V) in each bin.
Figure \ref{lohistred} shows the distribution of the optical luminosities before and after de-reddening. The $\lo$ distribution of the reddened AGN is significantly different (i.e. lower luminosity) from that of the total sample. This would suggest that extinction is more important for lower luminosity AGN (see \citealt{2004ASPC..311...61G}). 
The average shift induced by the correction for the intrinsic extinction in the 10\% of the total sample is $\langle\Delta \Log \lo \rangle=0.28 \pm 0.07$. 
\par
X-ray absorption is generally negligible in Type 1 AGN; however, it is known that a fraction of the order of 10\% of broad-line AGN may be obscured by column densities up to 10$^{22}$ cm$^{-2}$ (see \citealt{mainieri07}).
Unabsorbed X-ray fluxes can be computed if the absorption column density ($N_H$) is known, which is not the case for most of the sources in our sample. Hardness ratios may be used instead, but they are almost insensitive to column densities of the order of 10$^{22}$cm$^{-2}$ or slightly higher at the average redshifts of the XMM-COSMOS sources, and they tend to over-estimate $N_H$ (e.g. \citealt{2004A&A...421..491P}). In order to quantify the average impact of X-ray absorption on the $\alphaox$ distribution and bolometric corrections, we have assumed that 10\% of the sources in our sample are obscured by a column density of $10^{22}$ cm$^{-2}$. We note that this assumption is likely to overestimate X-ray absorption in Type 1 AGN (see Fig.~13 in \citealt{mainieri07}).
By correcting X-ray monocromatic fluxes at 2 keV for randomly chosen 10\% of the sources, the unobscured X-ray fluxes are 10\% higher. The shift induced by this correction in the 10\% of the total sample is $\langle\Delta \Log \lx \rangle=0.04$.
\par
Broad absorption-line quasars (BAL QSOs) are known to be X-ray obscured (e.g., \citealt{green1995ApJ...450...51G}, \citealt{1999ApJ...519..549G}, \citealt{brandt2000}), and are not included in previous studies of optically selected samples because they can cause an artificial steepening of the $\alphaox-\lo$ and $\lo-\lx$ correlations.
Due to the lack of a systematic analysis of the optical spectra of broad-line AGN in the zCOSMOS and IMACS spectroscopic surveys and the inclusion of photometric Type 1 AGN, we do not have an estimate of the BAL fraction in our sample.
Assuming they constitute about 10--15\% of the entire quasar population (e.g. R03, \citealt{2003AJ....125.1784H}), BAL QSO may be numerous among red sources in Fig.\ref{lohistred} and/or X-ray obscured AGN.
Therefore, we expect that the effects of dust reddening and X-ray absorption considered and quantified in the previous paragraphs take into account, at least in a statistical sense, the BAL QSO contamination. Moreover, the considered fraction of BAL QSOs comes from optically selected samples, so that the same fraction should be smaller (and possibly redshift dependent) for soft X-ray selected samples.
\par
Applying these corrections for absorption in the optical and in the X-ray, we find that the slopes of the $\lo-\lx$ and $\alphaox-\lo$ relations become $\beta_{corr}=0.782\pm0.021$ (OLS bisector algorithm) and $0.147\pm0.008$. Both these ``corrected" slopes are within one sigma from those derived with no correction for absorption (see eq. [\ref{lo_lx_bisector}] and eq. [\ref{aoxlo_loind}]). The slight steepening of the $\lo-\lx$ relation is due to the fact that the fraction of optically extincted AGN is higher at lower optical luminosity (see Fig. \ref{lohistred}). The corrections for absorption increase the mean value of $\alphaox$ by only 0.01. We therefore conclude that the absorption corrections do not change significantly our overall results.

\subsection{Host-Galaxy Contamination}
\label{Host Galaxy Contamination}

\begin{figure}                  
 \resizebox{\hsize}{!}{\includegraphics{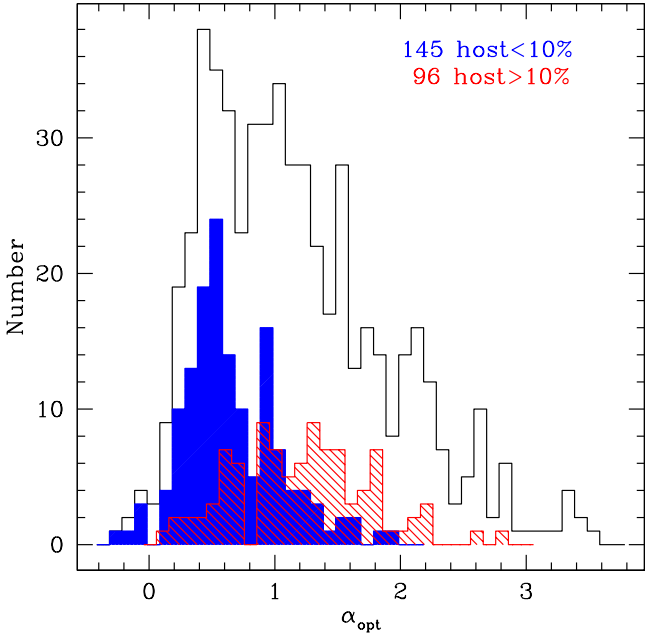}}
 \caption{Distribution of the optical index for the total sample (\textit{open histogram}), for the 145 AGN which have a host-galaxy contribution less than 10\% (\textit{filled histogram}) and for the 96 AGN which have a host-galaxy contribution greater than 10\% (\textit{hatched histogram}).}
 \label{histao}        
\end{figure}          

In order to estimate the possible effect of the contribution of the host-galaxy stellar light to the $\Alambda$ luminosity, we have considered a subsample of 241 sources for which we have a host-galaxy contribution computed subtracting a point-spread-function scaled to a central 4 pixel aperture, and measuring the extended flux in ACS F814W (\citealt{2004ApJ...614..568J}, \citealt{jahnke09}): 145 sources have an extended host-galaxy contribution of less than 10\%, while for 96 AGN the contamination of the host-galaxy light is greater than 10\%. This is a robust, model independent measure to determine whether substantial extended flux is present or not. A useful parameter that we can compute for the total sample is the optical spectral index, $\alphao$. By linking it with the morphological information we can try to quantify the host-galaxy contribution for each AGN.

Following \citet{vandenberk2001}, we compute the optical spectral index using the monochromatic luminosities at $2000\text{\AA}$ and $5000\text{\AA}$
\begin{equation}
     \alphao=-2.5 \,\Log \frac{L_{2000\text{\AA}}}{L_{5000\text{\AA}}}.
\end{equation}
Figure \ref{histao} shows the distribution of $\alphao$ for the total sample, for the 145 sources for which the stellar light contribution to the total flux is less than $\sim10\%$ and for the 96 sources for which this contribution is greater than 10\%.
These two samples have significantly different distributions of $\alphao$. Unresolved quasars have an average $\alphao$ of $\sim0.5$, which is consistent with previous results from the SDSS samples, while the sample with a significant galaxy contribution has a range of $\alphao$ from 0 to $\sim3$ (typical of a spiral Sc template) with an average of $\sim 1.2$.

We then tried to estimate the fraction of galaxy light at $\Alambda$ assuming a typical slope $\alphao=0.5$ for the nucleus component and $\alphao=3$ for the galaxies. We further assume that the monochromatic luminosity in the range $2000-5000\text{\AA}$ is due to the contribution of two power laws:
\begin{equation}
     L_\nu = A \nu^{-0.5} + G \nu^{-3},
\end{equation}
where $A$ and $G$ are the normalization constants we want to estimate.
With this assumption the observed $\alphao$ is a function only of the ratio between $A$ and $G$, therefore, 
at any given $\alphao$ value, the expected relative contribution of the nuclear and host-galaxy emission can be computed.
We divided the total sample in 5 bins of $\alphao$, with $\Delta \alphao \sim 0.5$, and corrected the observed $\lo$ using an average value of $A/G$ in each bin. The shift in the monochromatic luminosities at $\Alambda$ induced using this correction is $\langle\Delta \Log \lo\rangle=-0.15\pm 0.01$, with a dispersion of 0.21.
\par
Applying this correction, we find that the slopes of the $\lo-\lx$ and $\alphaox-\lo$ become $0.660\pm0.022$ (OLS bisector algorithm) and $0.197\pm0.011$, respectively. Both these slopes corrected for the possible host-galaxy contribution are different at about $3\sigma$ from those derived with no correction (see eq. [\ref{lo_lx_bisector}] and [\ref{aoxlo_loind}]). In this case, the flattening of the $\lo-\lx$ relation is due to the fact that $\alphao$ tends to be steeper (i.e. with a possibly higher contribution from the host-galaxy) for lower luminosity AGN. This correction decreases the mean value of $\alphaox$ by $\sim0.06$.

At face value, the possible effects of host-galaxy contribution estimated above are not negligible. However, we stress that these estimates have to be taken as upper limits, for at least two reasons. First, in our derivation of the constants $A$ and $G$ we have assumed that all AGN have the same intrinsic $\alphao$; in presence of a dispersion of $\alphao$ distribution of AGN, the contribution from the host-galaxy would be smaller than that we have derived. Secondly, the same would happen also if, as it is likely to be the case, both effects (extinction and host-galaxy contribution) are at work at the same time.

\section{Discussion}
\label{Discussion}

\subsection {Sample biases and systematics}

We used the COSMOS multi-wavelength database to build a large X-ray selected sample of Type 1 AGN and study their optical to X-ray properties. The sample size is comparable to that of optically selected samples previously reported in the literature (e.g., \citealt{2008ApJ...685..773G}), but to our knowledge this is the largest complete X-ray selected sample for which the study of the $\alphaox$ distribution is performed.
In order to keep the selection criteria as clean as possible and to cope with the lack of spectroscopic information at faint optical magnitudes, the Type 1 classification is based on both spectroscopic and photometric redshifts.
This choice guarantees a well defined and relatively simple selection function, but it is not completely free from other biases.
In particular, contamination from photometrically misclassified Type 1 AGN cannot be excluded (see Section \ref{The Broad Line AGN Sample}).
Other possible biases and systematics include the contribution of the effect of dust reddening and gas absorption at UV and X-ray frequencies and the host-galaxy light to the $\Alambda$ nuclear flux. Even though we tried to estimate and quantify the impact of the above mentioned biases (see Section \ref{Intrinsic Extinction} and \ref{Host Galaxy Contamination}), some residual contamination due to either one or more of these biases cannot be completely excluded. However, we are confident that the most important results of our analysis, and especially those concerning the average properties of the total sample, are not significantly affected.

The correlations between optical and X-ray luminosities obtained from the analysis of the spectroscopic sample are recovered, if only photo-z are considered.
Even though the best fit parameters of the correlations are slightly different, the inclusion of photo-z sources allow us to extend the study of optical and X-ray properties to much lower luminosities.

\subsection{Constraints on the X-ray emission models}

The optical to X-ray spectral index $\alphaox$ connects the two portions of the AGN broad-band spectrum dominated by the accretion power and
thus it is expected to be a reliable tracer of the accretion properties and, in particular, of the relation between disk emission, peaking in the UV, and coronal X-ray emission. Previous studies have shown a non linear dependence of the X-ray versus UV luminosity, which implies a correlation between the optical--UV to X-ray luminosity ratio on monochromatic luminosity and/or redshift. Based on extensive analysis of large samples of optically selected AGN (i.e. \citealt{vignali03}, \citealt{strateva05}, S06, J07) it is concluded that $\alphaox$ is primary dependent on optical luminosity at $\Alambda$.
The results presented in this paper confirm and extend the previous findings to a large sample of X-ray selected AGN, suggesting that band selection does not significantly modify the current observational picture.
The observed $\lo-\lx$ (or $\alphaox-\lo$) correlation implies that more optical luminous AGN emit less X-rays per unit UV luminosity than less luminous AGN.
The predicted UV to X-ray luminosity ratio depends on the hot corona covering factor, optical depth and electron temperatures.
The optical--UV and X-ray emission of radio-quiet AGN can be explained by the interplay between hot electrons in a coronal gas and a colder
accretion flow. Soft photons from the accretion disk are Comptonized by hot electrons and lead to the formation of a power law spectrum in the hard X-rays accompanied by a high energy cut-off at the electrons' temperature (\citealt{1991ApJ...380L..51H, 1993ApJ...413..507H}).
If only a fraction of the accretion power is released in the hot phase, as in the patchy corona model (\citealt{1994ApJ...432L..95H}), the $L_{\rm UV}/L_{\rm X}$ ratio is higher than the value computed using a model with more uniform corona. The observed correlation suggest that disk-corona parameters are depending on UV luminosity.

\subsection {Using $\alphaox$ as a bolometric correction estimator}

An accurate determination of bolometric quantities is essential to compute AGN accretion rates and, more in general,
to all the arguments related to the accretion onto SMBH at large. These parameters require to sample the AGN SED over
a broad range of wavelengths from near-infrared to hard X-rays. Owing to the excellent quality of the multi-wavelength COSMOS database, this is becoming possible for large AGN samples (Elvis et al. 2009). Because the optical--UV to X-ray portion of the spectrum contains about 60\% of the total bolometric luminosity, we have investigated to what extent the UV to X-ray luminosity ratio can be considered a reliable proxy of the bolometric correction.

We computed bolometric corrections from the 2--10 keV energy range for all the objects in our sample with detection in both soft and hard bands.
Given the strong luminosity dependence of $\alphaox$ upon UV luminosity which is close to the SED maximum, it is not surprising
to find a significant correlation between $\alphaox$ and $\kbol$.
However, it is important to note that the correlation, best-fitted by a second order polynomial, has a small dispersion around the best-fit (see Fig. \ref{kbolaox}).
The existence of such a tight relation between these two parameters suggests that the AGN bolometric output is well traced by $\alphaox$ over a broad range of redshifts and luminosities.
Moreover, it offers the opportunity to estimate reliable bolometric corrections, for large AGN samples, provided that rest-frame $2500\text{\AA}$ and 2 keV luminosities are known.

\subsection{Bolometric corrections and Eddington ratios}
Accretion rates are then estimated using the previously computed bolometric corrections.
A correlation is found between the 2--10 keV bolometric correction and the Eddington ratio for a sizable subsample of 150 objects (see Fig. \ref{kboleddcheck}), for which black hole masses are computed from the broad emission lines FWHM (see Merloni et al. 2009, \citealt{trump09} for a detailed discussion).
The presence of this correlation was originally suggested by \citet{vasudevanfabian07} using broad-band observations of nearby Seyfert galaxies and confirmed in a subsequent paper (VF09) using simultaneous optical-UV and X-ray data. A similar trend is reported by \citet{bianchi09} from a multi-wavelength analysis of archival XMM-\textit{Newton} observations.
Although we do not have simultaneous data or $\MBH$ estimate from reverberation mapping, our findings are consistent with the $\kbol-\lambdaEdd$ relation by VF09.
This correlation may, in principle, be induced by the fact that both parameters depend on the bolometric luminosity. However, if we look for the effect of the bolometric luminosity using the Kendall-$\tau$ partial correlation analysis, we found that the correlation is still significant. 
This suggests that the systematic effects introduced by the presence of bolometric luminosity on both axes are not relevant (see also Section 5.1.4 in VF07). 

A fairly significant correlation, albeit with a large scatter, is also found between $\alphaox$ and the Eddington ratio.
In the framework of the disk-corona models discussed above, sources with steep $\alphaox$ are explained by a patchy corona where the number of blobs or their covering factor decreases for increasing values of $\alphaox$.
For a given BH mass, bright UV emission is due to a high accretion rate. If in highly accreting sources most of the power is dissipated in the disk, rather than in the hot corona, then a correlation between $\alphaox$ and Eddington ratio is expected.

\section{Summary and conclusions}
\label{Summary and conclusions}

In this paper we have presented the analysis of the UV to X-ray properties of 545 radio-quiet X-ray selected Type 1 AGN, in  the multi-wavelength COSMOS survey.
The full data-set covers a large range of redshifts ($0.04<\z<4.25$) and X-ray luminosities ($40.6 \leq \Log \Lhard \leq 45.3$). It is mostly composed by spectroscopically selected Type 1 AGN ($322$ sources, $\sim 60\%$ of the total sample), with the addition of $223$ Type 1 AGN classified on the basis of the best-fitting SED procedure provided by S09. We constructed single SEDs for the full sample and, from these, we computed optical and X-ray rest-frame luminosities at $\Alambda$ and 2 keV, respectively; we also investigated the dependence of $\alphaox$ upon redshift, $\lx$ and $\lo$ using the fully parametric EM (estimated and maximized) regression algorithm and Kendall-$\tau$ partial correlation analysis. Moreover, we estimated bolometric correction and bolometric luminosities for a subsample of 343 AGN for which we have detections in both soft and hard bands. Our principal results are the following:
\begin{enumerate}
 \item We confirm the $\lx-\lo$ correlation parametrized by $\lx \propto \lo^\beta$, where $\beta=0.760\pm0.022$. The best-fit slope agrees with previous studies based on optically selected samples, which found a value of $\beta$ inconsistent with unity.

 \item The mean value of $\alphaox$ for the full X-ray selected sample is $\langle\alphaox\rangle\sim 1.37 \pm 0.01$ with a dispersion around the mean of 0.18. We confirm the correlation between $\alphaox$ and $\lo$ at the $17\,\sigma$ significance level. The correlation becomes stronger if we take into account the effect of redshift using the partial correlation analysis ($\sim 21\,\sigma$).
The slope of our the best-fit relation between $\alphaox$ and $\lo$ is consistent at least within $\sim 1.6\sigma$ with that obtained for an optically selected sample.

 \item From the EM regression method, we find a weak correlation between $\alphaox$ and redshift; if the effect of $\lo$ is taken into account, the correlation disappears.

 \item We do not find any significant correlation between $\alphaox$ and $\lx$ ($\sim 1.5\,\sigma$ significance level if we take into account the effect of the optical luminosity).

 \item We found a tight correlation between the $\alphaox$ values and the hard X-ray bolometric correction values. The $\alphaox-\kbol$ relation could be used as a practical tool to provide an accurate estimate ($\sim 20\%$ at $1\sigma$) of the bolometric correction using only the $\alphaox$ value.

 \item We found a correlation for both $\alphaox$ and $\kbol$ with Eddington ratio. Our results for the $\kbol-\lambdaEdd$ relation and for $\alphaox-\lambdaEdd$ suggest that there is a connection between the broad-band emission, mostly in the UV, and the Eddington ratio, which is directly linked to the ratio between mass accretion rate, $\dot M_{\rm acc}$, and Eddington accretion rate, $\dot M_{\rm Edd}$.

 \item We have estimated the possible effects of intrinsic absorption and/or contribution of the host-galaxy on the $\lo-\lx$ and $\alphaox-\lo$ relations. If the main reason for the red optical-UV SEDs in our sample were intrinsic extinction, the resulting $\lo-\lx$ relation would be slightly steeper ($\beta=0.782\pm0.021$); if, instead, the host-galaxy contribution were important in determining the red optical-UV SEDs, the resulting $\lo-\lx$ relation would be flatter ($\beta=0.660\pm0.022$). Neither effect can produce a linear correlation ($\beta=1$).

\end{enumerate}

\begin{acknowledgements}
In Italy, the XMM-COSMOS project is supported by PRIN/MIUR under grant 2006-02-5203, ASI-INAF grants I/023/05/00, I/088/06 and ASI/COFIS/WP3110
I/026/07/0. In Germany the XMM-\textit{Newton} project is supported by the Bundesministerium f\"{u}r Wirtshaft und Techologie/Deutsches Zentrum f\"{u}r Luft und Raumfahrt and the Max-Planck society. The entire COSMOS collaboration is gratefully acknowledged.
\end{acknowledgements}

\bibliographystyle{aa}
\bibliography{bibl}


\end{document}